\documentclass[%
 reprint,
superscriptaddress,
showpacs,preprintnumbers,
 amsmath,amssymb,
 aps,
]{revtex4-1}

\usepackage{graphicx}
\usepackage{dcolumn}
\usepackage{bm}

\begin{document}

\renewcommand{\textfraction}{0.00000000001}
\renewcommand{\floatpagefraction}{1.0}

\preprint{APS/123-QED}

\title{\boldmath
Target and beam-target asymmetries for the $\gamma p\to \pi^0\pi^0 p$ reaction}

\author{S.~Garni}
\affiliation{Department of Physics, University of Basel, Ch-4056 Basel, Switzerland}
\author{V.~L.~Kashevarov}\email[]{Electronic address: kashev@kph.uni-mainz.de}
\affiliation{Institut f\"ur Kernphysik, University of Mainz, D-55099 Mainz, Germany}
\affiliation{Joint Institute for Nuclear Research, 141980 Dubna, Russia}
\author{A.~Fix}\email[]{Electronic address: fix@mph.phtd.tpu.ru}
\affiliation{Tomsk Polytechnic University, Tomsk, Russia}
\author{S.~Abt}
\affiliation{Department of Physics, University of Basel, Ch-4056 Basel, Switzerland}
\author{F.~Afzal}
\affiliation{Helmholtz-Institut f\"ur Strahlen- und Kernphysik, University Bonn, D-53115 Bonn, Germany}
\author{P.~Aguar Bartolome}
\affiliation{Institut f\"ur Kernphysik, University of Mainz, D-55099 Mainz, Germany}
\author{Z.~Ahmed}
\affiliation{University of Regina, Regina, SK S4S-0A2 Canada}
\author{J.~Ahrens}
\affiliation{Institut f\"ur Kernphysik, University of Mainz, D-55099 Mainz, Germany}
\author{J.R.M.~Annand}
\affiliation{SUPA School of Physics and Astronomy, University of Glasgow, Glasgow, G12 8QQ, UK}
\author{H.J.~Arends}
\affiliation{Institut f\"ur Kernphysik, University of Mainz, D-55099 Mainz, Germany}
\author{M.~Bashkanov}
\affiliation{SUPA School of Physics, University of Edinburgh, Edinburgh EEH9 3JZ, UK}
\author{R.~Beck}
\affiliation{Helmholtz-Institut f\"ur Strahlen- und Kernphysik, University Bonn, D-53115 Bonn, Germany}
\author{M.~Biroth}
\affiliation{Institut f\"ur Kernphysik, University of Mainz, D-55099 Mainz, Germany}
\author{N.~Borisov}
\affiliation{Joint Institute for Nuclear Research, 141980 Dubna, Russia}
\author{A.~Braghieri}
\affiliation{INFN Sezione di Pavia, I-27100 Pavia, Pavia, Italy}
\author{W.J.~Briscoe}
\affiliation{Center for Nuclear Studies, The George Washington University, Washington, DC 20052, USA}
\author{S.~Cherepnya}
\affiliation{Lebedev Physical Institute, RU-119991 Moscow, Russia}
\author{F.~Cividini}
\affiliation{Institut f\"ur Kernphysik, University of Mainz, D-55099 Mainz, Germany}
\author{C.~Collicott}
\affiliation{Department of Astronomy and Physics, Saint Mary's University, E4L1E6 Halifax, Canada}
\author{S.~Costanza}\altaffiliation{Also at: Dipartimento di Fisica, Universit\`a di Pavia, I-27100 Pavia, Italy}
\affiliation{INFN Sezione di Pavia, I-27100 Pavia, Pavia, Italy}
\author{A.~Denig}
\affiliation{Institut f\"ur Kernphysik, University of Mainz, D-55099 Mainz, Germany}
\author{A.S.~Dolzhikov}
\affiliation{Joint Institute for Nuclear Research, 141980 Dubna, Russia}
\author{E.J.~Downie}
\affiliation{Center for Nuclear Studies, The George Washington University, Washington, DC 20052, USA}
\author{P.~Drexler}
\affiliation{Institut f\"ur Kernphysik, University of Mainz, D-55099 Mainz, Germany}
\affiliation{II. Physikalisches Institut, University of Giessen, D-35392 Giessen, Germany}
\author{L.V.~Fil'kov}
\affiliation{Lebedev Physical Institute, RU-119991 Moscow, Russia}
\author{S.~Gardner}
\affiliation{SUPA School of Physics and Astronomy, University of Glasgow, Glasgow, G12 8QQ, UK}
\author{D.I.~Glazier}
\affiliation{SUPA School of Physics and Astronomy, University of Glasgow, Glasgow, G12 8QQ, UK}
\affiliation{SUPA School of Physics, University of Edinburgh, Edinburgh EEH9 3JZ, UK}
\author{I.~Gorodnov}
\affiliation{Joint Institute for Nuclear Research, 141980 Dubna, Russia}
\author{W.~Gradl}
\affiliation{Institut f\"ur Kernphysik, University of Mainz, D-55099 Mainz, Germany}
\author{M.~G{\"u}nther}
\affiliation{Department of Physics, University of Basel, Ch-4056 Basel, Switzerland}
\author{D.~Gurevich}
\affiliation{Institute for Nuclear Research, RU-125047 Moscow, Russia}
\author{L. Heijkenskj{\"o}ld}
\affiliation{Institut f\"ur Kernphysik, University of Mainz, D-55099 Mainz, Germany}
\author{D.~Hornidge}
\affiliation{Mount Allison University, Sackville, New Brunswick E4L1E6, Canada}
\author{G.M.~Huber}
\affiliation{University of Regina, Regina, SK S4S-0A2 Canada}
\author{A.~K{\"a}ser}
\affiliation{Department of Physics, University of Basel, Ch-4056 Basel, Switzerland}
\author{S.~Kay}
\affiliation{SUPA School of Physics, University of Edinburgh, Edinburgh EEH9 3JZ, UK}
\author{I.~Keshelashvili}\altaffiliation{Present adaress: Institut f\"ur Kernphysik, FZ J\"ulich, 52425 J\"ulich, Germany}
\affiliation{Department of Physics, University of Basel, Ch-4056 Basel, Switzerland}
\author{R.~Kondratiev}
\affiliation{Institute for Nuclear Research, RU-125047 Moscow, Russia}
\author{M.~Korolija}
\affiliation{Rudjer Boskovic Institute, HR-10000 Zagreb, Croatia}
\author{B.~Krusche}
\affiliation{Department of Physics, University of Basel, Ch-4056 Basel, Switzerland}
\author{A.~Lazarev}
\affiliation{Joint Institute for Nuclear Research, 141980 Dubna, Russia}
\author{V.~Lisin}
\affiliation{Institute for Nuclear Research, RU-125047 Moscow, Russia}
\author{K.~Livingston}
\affiliation{SUPA School of Physics and Astronomy, University of Glasgow, Glasgow, G12 8QQ, UK}
\author{S.~Lutterer}
\affiliation{Department of Physics, University of Basel, Ch-4056 Basel, Switzerland}
\author{I.J.D.~MacGregor}
\affiliation{SUPA School of Physics and Astronomy, University of Glasgow, Glasgow, G12 8QQ, UK}
\author{R.~Macrae}
\affiliation{SUPA School of Physics and Astronomy, University of Glasgow, Glasgow, G12 8QQ, UK}
\author{Y.~Maghrbi}\altaffiliation{Present address: College of Engineering and Technology,\\
American University of the Middle East, Kuwait}
\affiliation{Department of Physics, University of Basel, Ch-4056 Basel, Switzerland}
\author{D.M.~Manley}
\affiliation{Kent State University, Kent, Ohio 44242, USA}
\author{P.P.~Martel}
\affiliation{Institut f\"ur Kernphysik, University of Mainz, D-55099 Mainz, Germany}
\affiliation{Mount Allison University, Sackville, New Brunswick E4L3B5, Canada}
\author{J.C.~McGeorge}
\affiliation{SUPA School of Physics and Astronomy, University of Glasgow, Glasgow, G12 8QQ, UK}
\author{D.G.~Middleton}
\affiliation{Mount Allison University, Sackville, New Brunswick E4L3B5, Canada}
\author{R.~Miskimen}
\affiliation{University of Massachusetts, Amherst, Massachusetts 01003, USA}
\author{E.~Mornacchi}
\affiliation{Institut f\"ur Kernphysik, University of Mainz, D-55099 Mainz, Germany}
\author{C.~Mullen}
\affiliation{SUPA School of Physics and Astronomy, University of Glasgow, Glasgow, G12 8QQ, UK}
\author{A.~Mushkarenkov}
\affiliation{INFN Sezione di Pavia, I-27100 Pavia, Pavia, Italy}
\affiliation{University of Massachusetts, Amherst, Massachusetts 01003, USA}
\author{A.~Neganov}
\affiliation{Joint Institute for Nuclear Research, 141980 Dubna, Russia}
\author{A.~Neiser}
\affiliation{Institut f\"ur Kernphysik, University of Mainz, D-55099 Mainz, Germany}
\author{M.~Oberle}
\affiliation{Department of Physics, University of Basel, Ch-4056 Basel, Switzerland}
\author{M.~Ostrick}
\affiliation{Institut f\"ur Kernphysik, University of Mainz, D-55099 Mainz, Germany}
\author{P.B.~Otte}
\affiliation{Institut f\"ur Kernphysik, University of Mainz, D-55099 Mainz, Germany}
\author{B.~Oussena}
\affiliation{Institut f\"ur Kernphysik, University of Mainz, D-55099 Mainz, Germany}
\affiliation{Center for Nuclear Studies, The George Washington University, Washington, DC 20052, USA}
\author{D.~Paudyal}
\affiliation{University of Regina, Regina, SK S4S-0A2 Canada}
\author{P.~Pedroni}
\affiliation{INFN Sezione di Pavia, I-27100 Pavia, Pavia, Italy}
\author{F.~Pheron}
\affiliation{Department of Physics, University of Basel, Ch-4056 Basel, Switzerland}
\author{A.~Polonski}
\affiliation{Institute for Nuclear Research, RU-125047 Moscow, Russia}
\author{A.~Powell}
\affiliation{SUPA School of Physics and Astronomy, University of Glasgow, Glasgow, G12 8QQ, UK}
\author{S.N.~Prakhov}
\affiliation{University of California Los Angeles, Los Angeles, California 90095-1547, USA}
\author{G.~Ron}
\affiliation{Racah Institute of Physics, Hebrew University of Jerusalem, Jerusalem 91904, Israel}
\author{T.~Rostomyan}\altaffiliation{Present address: Department of Physics and Astronomy, Rutgers University,
Piscataway, New Jersey, 08854-8019}
\affiliation{Department of Physics, University of Basel, Ch-4056 Basel, Switzerland}
\author{A.~Sarty}
\affiliation{Department of Astronomy and Physics, Saint Mary's University, E4L1E6 Halifax, Canada}
\author{C.~Sfienti}
\affiliation{Institut f\"ur Kernphysik, University of Mainz, D-55099 Mainz, Germany}
\author{V.~Sokhoyan}
\affiliation{Institut f\"ur Kernphysik, University of Mainz, D-55099 Mainz, Germany}
\author{K.~Spieker}
\affiliation{Helmholtz-Institut f\"ur Strahlen- und Kernphysik, University Bonn, D-53115 Bonn, Germany}
\author{O.~Steffen}
\affiliation{Institut f\"ur Kernphysik, University of Mainz, D-55099 Mainz, Germany}
\author{I.I.~Strakovsky}
\affiliation{Center for Nuclear Studies, The George Washington University, Washington, DC 20052, USA}
\author{I.~Supek}
\affiliation{Rudjer Boskovic Institute, HR-10000 Zagreb, Croatia}
\author{A.~Thiel}
\affiliation{Helmholtz-Institut f\"ur Strahlen- und Kernphysik, University Bonn, D-53115 Bonn, Germany}
\author{M.~Thiel}
\affiliation{Institut f\"ur Kernphysik, University of Mainz, D-55099 Mainz, Germany}
\author{A.~Thomas}
\affiliation{Institut f\"ur Kernphysik, University of Mainz, D-55099 Mainz, Germany}
\author{M.~Unverzagt}
\affiliation{Institut f\"ur Kernphysik, University of Mainz, D-55099 Mainz, Germany}
\author{Yu.A.~Usov}
\affiliation{Joint Institute for Nuclear Research, 141980 Dubna, Russia}
\author{S.~Wagner}
\affiliation{Institut f\"ur Kernphysik, University of Mainz, D-55099 Mainz, Germany}
\author{N.K.~Walford}
\affiliation{Department of Physics, University of Basel, Ch-4056 Basel, Switzerland}
\author{D.P.~Watts}
\affiliation{SUPA School of Physics, University of Edinburgh, Edinburgh EEH9 3JZ, UK}
\author{J.~Wettig}
\affiliation{Institut f\"ur Kernphysik, University of Mainz, D-55099 Mainz, Germany}
\author{M.~Wolfes}
\affiliation{Institut f\"ur Kernphysik, University of Mainz, D-55099 Mainz, Germany}
\author{L.A.~Zana}
\affiliation{SUPA School of Physics, University of Edinburgh, Edinburgh EEH9 3JZ, UK}
\author{F. Zehr}
\affiliation{Department of Physics, University of Basel, Ch-4056 Basel, Switzerland}
\collaboration{A2 Collaboration}

\date{today}

\begin{abstract}
\begin{description}
\item[Background]
{Photoproduction of pion pairs allows the study of sequential decays of nucleon resonances
via excited intermediate states. Such decays are important for complex states involving 
more than one quark excitation which de-excite in a two-step process. 
However, the analysis of multi-meson final states is difficult and generally relies
on measurement of polarization observables.}
\item[Purpose]
{Experimental measurement and analysis of target and beam-target polarization observables 
of the $\gamma p\rightarrow \pi^0\pi^0$ reaction.} 
\item[Methods]
{Target (single) and beam-target (double) polarization asymmetries were investigated as a function of
several parameters. The experiments were performed at the Mainz Microtron (MAMI) laboratory using
circularly polarized photon beams and transversally polarized solid-butanol targets. The reaction
products were analyzed with a near 4$\pi$ solid-angle electromagnetic calorimeter composed of the
Crystal Ball and TAPS detectors.}
\item[Results]
{The polarization observables studied were $P_y$ (unpolarized beam, target polarized in the $y$ 
direction) and $P_x^{\odot}$ (circularly polarized beam, target polarized in the $x$ direction), 
which are similar to $T$ (target asymmetry) and $F$ (beam-target asymmetry)
for single meson production. The asymmetries were analyzed with three independent methods,
revealing systematic uncertainties. Some results are also given for the asymmetries $P_x$ and
$P_y^{\odot}$ which contribute only for three-body final states.}
\item[Conclusions]
{The measured observables allow some general conclusions to be drawn about the resonance content 
of the reaction amplitude. The $3/2^-$ partial wave shows a clear resonant behavior, attributed in 
the second resonance region to the sequential decay of the $N(1520)3/2^-$. The behavior of the  
$1/2^-$, $3/2^+$ partial waves is much smoother and the origin of a $3/2^+$ component at low energies 
is not well understood. The new data are important for future analyses of the partial wave structure 
of the $\pi^0\pi^0$ photoproduction amplitude. However, further experimental results for the 
isospin dependence and the helicity decomposition of the reaction are needed.}
\end{description}
\end{abstract}


\maketitle

\section{Introduction}\label{sec:introduction}

Photoproduction of pairs of pseudoscalar mesons from nucleons, in particular $\pi\pi$
\cite{Braghieri_95,Haerter_97,Krusche_99,Wolf_00,Kleber_00,Assafiri_03,Kotulla_04,Ahrens_05,Ajaka_07,Sarantsev_08,Thoma_08,Krambrich_09,Zehr_12,Kashevarov_12,Oberle_13,Dieterle_15,Sokhoyan_15a,Sokhoyan_15b,Thiel_15,Zabrodin_97,Zabrodin_99,Langgaertner_01,Strauch_05,Oberle_14}
and $\pi\eta$ pairs (see \cite{Krusche_15} and Refs. therein for a summary and
\cite{Kaeser_15,Kaeser_16,Sokhoyan_18,Kaeser_18} for most recent results), has been studied
intensively during the last two decades. Many data for angular, energy, and invariant mass
distributions of the unpolarized cross sections as well as for some polarization observables
have been collected for proton and quasi-free neutron targets and for the production of both
charged and neutral mesons.

The purpose of many of these experiments is to extract the partial wave structure of the reaction
amplitudes in order to study the contributions of nucleon resonances in the presence of
non-resonant backgrounds. The main motivation is that excited nucleon states,
due to their internal structure, can have much larger decay probabilities for sequential
decays, involving intermediate states, than for direct decays to the nucleon ground state.
This is obvious for higher lying states that have two independent excited quark model oscillator 
modes and de-excite in a two-step process \cite{Thiel_15}.
However, already at fairly low excitation energies, e.g. for the $N(1520)3/2^-$ resonance,
significant branching ratios to sequential $\pi\pi$ decays via the $\Delta(1232)3/2^+$ state
have been observed \cite{Tejedor_96,Wolf_00}.

The most interesting but probably the least comprehensible channel in the production
of pion pairs is $\pi^0\pi^0 N$. Experimental data for this final state, in particular in the second 
and third resonance region are rapidly being accumulating. The investigation of this reaction,
which was not accessible to early experiments based on magnetic momentum analysis of charged
pions, was pioneered by the DAPHNE \cite{Braghieri_95} and TAPS \cite{Haerter_97} experiments
at MAMI with measurements of total cross sections. Subsequently many further results have
been reported for free proton and quasi-free neutron targets with and without polarization
degrees of freedom
\cite{Krusche_99,Wolf_00,Kleber_00,Assafiri_03,Kotulla_04,Ahrens_05,Ajaka_07,Sarantsev_08,Thoma_08,Krambrich_09,Zehr_12,Kashevarov_12,Oberle_13,Dieterle_15,Sokhoyan_15a,Sokhoyan_15b,Thiel_15}.

It has been frequently emphasized that this channel is especially convenient for studying
nucleon resonances, since in contrast to other channels with at least one charged
pion, only few reaction mechanisms contribute to $\pi^0\pi^0$ photoproduction. Other
contributions such as $\Delta$ Kroll-Ruderman term, pion-pole terms, $\rho$ photoproduction
etc. are forbidden. However, these peculiarities also represent a challenge to studying dynamical
properties of this channel using specific models. In absence of the well understood
Born terms, mentioned above, other background terms, whose nature is not well studied and which
are usually very model dependent, become relatively more important. This makes models for
$\pi^0\pi^0$ to some extent unreliable and is the main reason why significant qualitative
differences between the results of different analyses of $\pi^0\pi^0$ photoproduction still
exist. Interestingly, the model interpretation of this reaction is much less advanced than for the
$\pi^0\eta$ final state which is more strongly dominated by a few resonant terms
\cite{Gutz_14,Kaeser_16,Sokhoyan_18,Kaeser_18}.

Some analyses \cite{Tejedor_96,Sarantsev_08,Kashevarov_12} point to an important
role of the states with spin $J=3/2$ in $\pi^0\pi^0$ photoproduction, not only in the
vicinity of $N(1520)3/2^-$ but also at much lower energies. In Ref.\,\cite{Sarantsev_08}
this conclusion is based on a global fit to different reaction channels, such as $\gamma p\to\pi^0p$
$\pi^-p\to\pi^0\pi^0 p$, etc. This analysis claims evidence for a dominant contribution from the
partial wave with isospin $I=3/2$ and spin-parity $3/2^-$ in $\pi^0\pi^0$ photoproduction and assigns
the double-hump structure of the total cross section in the second and third resonance regions to
an interference of this wave (dominated by the $\Delta(1700)3/2^-$ resonance) with other contributions 
in particular from the $N(1520)3/2^-$ state. In a later analysis \cite{Sokhoyan_15b} also significant 
contributions from the $N(1680)5/2^-$ state to the third resonance peak were claimed.

In Ref.~\cite{Kashevarov_12} a partial wave expansion of the amplitude was analyzed. The measured
moments $W_{LM}$ of the angular distribution were fitted with this expansion. They represent the 
final state partial wave contributions with total momentum $J$ and projection $M$ on the normal to 
the plane spanned by the momenta of the final particles \cite{Fix_12}. It was shown that in the region 
up to $E_\gamma=800$ MeV, where only the lowest partial waves with $J\leq 3/2$ are expected to be 
important, the qualitative features of the moments $W_{LM}$ may unambiguously be interpreted in terms 
of the corresponding partial amplitudes. This allows rather firm conclusions about the dynamical 
content of the reaction to be drawn. In particular, it was found that the amplitude with $J^P=3/2^+$ 
may be as important as that with $J^P=3/2^-$. The problem behind this finding is that it is difficult 
to explain the origin of such a strong contribution of the $3/2^+$ wave.
Therefore, although the phenomenological model with a large $3/2^+$ wave contribution explains
rather well the observed angular and energy distributions in the reaction $\gamma p\to\pi^0\pi^0p$
\cite{Kashevarov_12} the mechanism of the $\pi^0\pi^0$ production is still not well understood.

\begin{figure}
\begin{center}
\resizebox{0.45\textwidth}{!}{%
\includegraphics{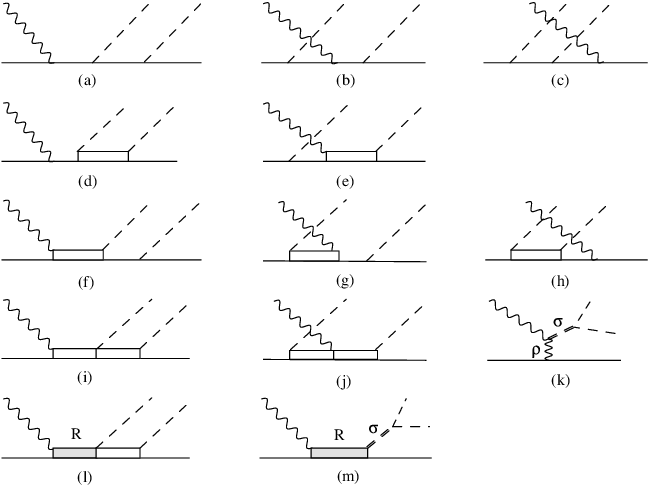}}
\caption{Diagrams for the reaction $\gamma p\to\pi^0\pi^0p$ used in the present work. The
empty rectangles represent the $\Delta(1232)$. Other resonances in the $s$ channel are
represented by shaded rectangles.}
\label{fig:diag}
\end{center}
\end{figure}

Better insight into details of the photoproduction of $\pi^0$ pairs requires the
measurement of further observables, which can better constrain the model analyses.
Of special interest are polarization observables because through interference terms
they are sensitive to contributions from small reaction amplitudes.
To date, besides the unpolarized cross section, experimental results for linear and
circular beam asymmetries \cite{Assafiri_03,Krambrich_09,Zehr_12,Oberle_13,Sokhoyan_15a,Sokhoyan_15b}
and beam-target asymmetries \cite{Ahrens_05} have already been collected and partially
been analyzed within different models \cite{Laget_96,Tejedor_96,Fix_05,Anisovich_05,Anisovich_12}.
It will certainly not be possible to do a `complete' experiment, leaving no ambiguities,
for the production of pseudoscalar meson pairs. This problem has not even been solved 
up to now for single $\pi^0$ photoproduction and there are significantly more degrees of 
freedom for pion pairs. While 8 observables need to be measured as a function of two 
kinematic variables for a unique solution for single pseudoscalar meson production 
\cite{Chiang_97}, pseudoscalar pairs require 8 observables to be measured as a function
of five kinematic variables just to fix the magnitudes of the amplitudes and require 
a total of 15 observables to also determine the phases \cite{Roberts_05}.

However, in some cases the partial wave content can be studied using a restricted 
number of observables while making some physically reasonable general assumptions about the
production mechanisms. Some of observables have already been measured and presented in 
earlier papers cited above. In the present work we expand the set of observables with 
the target and the beam-target asymmetries, which were not experimentally investigated before
now. An update of the model from \cite{Fix_05} was used for the interpretation of the data. 
The diagrams considered in this model are summarized in Fig.\,\ref{fig:diag}.
A new fit of the model to the recent data base for $\gamma N\rightarrow N\pi\pi$ has been 
made made.

\section{General formalism}

\begin{figure}
\begin{center}
\resizebox{0.45\textwidth}{!}{%
\includegraphics{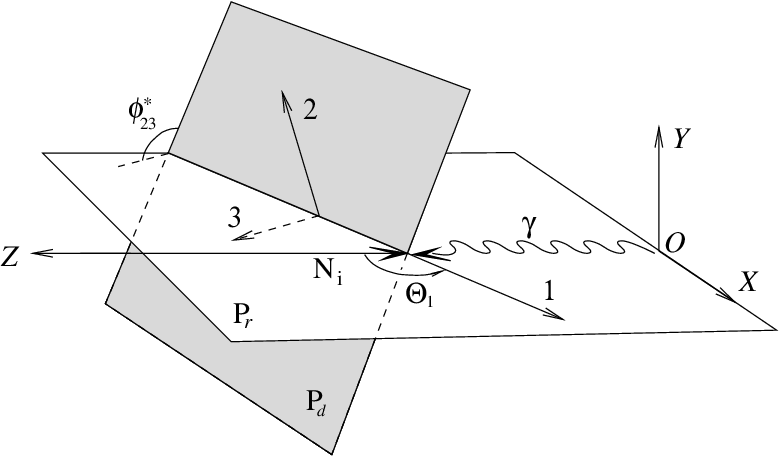}}
\caption{Definition of the coordinate system used in the present work. The azimuthal
angle $\phi_{23}^*$ is defined in the center of mass system of the particles 2 and 3 with
the $z$-axis opposite to the momentum of the particle 1 and $y$-axes parallel to $OY$. It
is equal to the angle between the reaction plane P$_r$ and the decay plane P$_d$.}
\label{fig:frame}
\end{center}
\end{figure}

The general formalism for the photoproduction of pseudoscalar meson pairs from nucleons was
developed in Refs.\,\cite{Roberts_05} and \cite{Fix_11}, where the formulae for different 
polarization observables were also presented. In the following we denote the final-state
particles as 1, 2, and 3 and consider the partition $1+(2\,3)$.
Because the two pions are identical particles, there are two independent sets of
variables, corresponding to the numbering $1+(2\,3)=p+(\pi\,\pi)$ and $\pi+(\pi\,p)$.
The coordinate system is shown in Fig.\,\ref{fig:frame}. The $Z$ axis is directed along the
photon momentum. The $X$ and $Y$ axes are chosen such that the momentum of the particle 1
has a positive $X$-projection and is orthogonal to the $Y$-axes. As independent kinematic
variables we choose the solid angle $\Omega_1=(\Theta_1,\Phi_1=0)$ of the particle 1, the
solid angle $\Omega_{23}^*=(\theta_{23}^*,\phi_{23}^*)$ of the particle 2 in the
center-of-mass of the pair $(2\,3)$ and the corresponding invariant mass $M_{23}$.

If the target nucleon is transversally polarized and the incident photon beam is
circularly polarized the cross section may be written in the form (see Eq.\,(57) of
Ref.\,\cite{Fix_11})
\begin{eqnarray}\label{eq1}
&&\frac{d\sigma}{d\Omega_1 dM_{23}d\Omega^*_{23}}= \frac{d\sigma_0}{d\Omega_1
dM_{23}d\Omega^*_{23}}\Big\{1+P_{\odot}
T^c_{00}\nonumber\\
&&\phantom{xxxx}-\frac{1}{\sqrt{2}}P_T\big[ T^0_{11}\cos\phi_s+S^0_{11}\sin\phi_s \\
&&\phantom{xxxx}+P_{\odot}(T^c_{11}\cos\phi_s+
S^c_{11}\sin\phi_s)\big]\Big\}\,,\nonumber
\end{eqnarray}
where the unpolarized differential cross section is denoted as $\sigma_0$. The values of
$P_T$ and $|P_{\odot}|$ describe the degree of nucleon polarization along the direction
determined by the angle $\Omega_s=(\theta_s=\frac{\pi}{2},\phi_s)$ and the degree of
photon circular polarization, respectively. The circular photon asymmetry $T^c_{00}$ was
already investigated in detail in Refs.\,\cite{Krambrich_09,Zehr_12,Oberle_13} and is
therefore excluded from the present study. As is seen from Eq.\,\ref{eq1}, for the totally
exclusive five-fold cross section, there are two independent transverse target asymmetries
($T^0_{11}$ and $S^0_{11}$) and two independent beam-target asymmetries ($T^c_{11}$ and
$S^c_{11}$). They are related to those introduced in Ref.\,\cite{Roberts_05} by
\begin{equation}\label{eq2}
T^0_{11}=-P_x\,,\quad S^0_{11}=-P_y\,,\quad  T^c_{11}=-P_x^\odot\,,\quad
S^c_{11}=-P_y^\odot\,.
\end{equation}

The corresponding semi-exclusive cross sections may be obtained easily from (\ref{eq1})
via appropriate integration. In particular, if one integrates over $\phi^*_{23}$, the
terms proportional to $T^c_{00}$, $T^0_{11}$, and $S^c_{11}$ vanish exactly, so that the
final expression reads
\begin{eqnarray}\label{eq3}
&&\frac{d\sigma}{d\Omega_1 dM_{23}}= \frac{d\sigma_0}{d\Omega_1 dM_{23}}\Big\{1-
\\
&&\phantom{xxxx}+\frac{1}{\sqrt{2}}P_T\big[\widetilde{S}^0_{11}\sin\phi_s
+P_{\odot}\widetilde{T}^c_{11}\cos\phi_s\big]\Big\}\,,\nonumber
\end{eqnarray}
where $\widetilde{T}^c_{11}$ and $\widetilde{S}^0_{11}$ are the corresponding partially
integrated (semi-exclusive) observables.

Table\,\ref{ta1} summarizes how the asymmetries discussed above can be separated by a
proper choice of the photon and proton polarization parameters. The observables $S^0_{11}$
and $T^c_{11}$ are similar to the ordinary $T$ and $F$ asymmetries used for single pion
photoproduction.

\begin{table}
\renewcommand{\arraystretch}{2.0}
\caption{Polarization observables measured in the present paper. In parentheses the
corresponding notations from Ref.\,\cite{Roberts_05} are given.} \label{ta1}
\begin{center}
\begin{tabular*}{8cm}
{@{\hspace{0.6cm}}c@{\hspace{0.6cm}}|@{\hspace{1.6cm}}c@{\hspace{1.6cm}}c}
\hline\noalign{\smallskip}
Beam & Target \\
\noalign{\smallskip}\hline\noalign{\smallskip}
     & x  &  y \\
\noalign{\smallskip}\hline\noalign{\smallskip}
$-$  & $T^0_{11}(P_x)$ & $S^0_{11}(P_y)$  \\
$c$  & $T^c_{11}(P_x^\odot)$ & $S^c_{11}(P_y^\odot)$  \\
\noalign{\smallskip}\hline
\end{tabular*}
\end{center}
\end{table}

\section{Experimental setup}
\label{sec:Setup}

The experiment was performed at the MAMI C accelerator in Mainz\,\cite{Kaiser_08} using the
Glasgow-Mainz tagged photon facility\,\cite{McGeorge_08,Hall_96,Anthony_91}.
The reaction $\gamma p\to \pi^0\pi^0 p$ was measured using the Crystal Ball
(CB)\,\cite{Starostin_01} as the central calorimeter and TAPS \,\cite{Novotny_91,Gabler_94}
as a forward detector. The CB detector is constructed as a sphere of 672 optically
insulated NaI(Tl) crystals, pointing toward the center of the sphere. The crystals are
arranged in two hemispheres that cover 93\% of the full solid angle. For charged-particle
identification a barrel of 24 scintillation counters, the Particle Identification Detector
(PID) \,\cite{Watts_05}, and two multiple wire proportional chambers
(MWPC) \cite{Audit_91} surrounded the target. The forward angular range $\theta =
5 - 20^\circ$ was covered by the TAPS calorimeter \,\cite{Novotny_91,Gabler_94}, that was
arranged in a plane consisting of 384 hexagonally shaped BaF$_2$ detectors. A 5-mm thick
plastic scintillator in front of each module allowed charged particles to be identified.
The solid angle covered by the Crystal Ball and TAPS detection system is nearly $97\%$ of
$4\pi$ sr. More details on the energy and angular resolution and particle identification
of the device are given in
Refs.\,\cite{Prakhov_09,McNicoll_10,Werthmueller_14,Witthauer_17,Dieterle_18}.

For the measurements discussed in this work longitudinally polarized electron beams with
energies of 1558~MeV and polarization degrees of $\approx$80\% were used. The polarization
of the electron beam was measured in special runs close to the electron source after the
Linac accelerator at beam energies of 3.65~MeV with Mott scattering and it was continuously
monitored with M$\o$ller scattering of the electrons from the ferromagnetic radiator foil
(Vacoflux50, 10 $\mu$m thickness). The longitudinal polarization of the electrons was
transferred to circular polarization of the photons according to \cite{Olsen_59}:
\begin{equation}
P_{\gamma}=P_{e^{-}}\cdot\frac{4x-x^{2}}{4-4x+3x^{2}}~,\label{eq:olsen}
\end{equation}
where $P_{e^{-}}$ and $P_{\gamma}$ are the degrees of polarization of the electrons and
the photons, respectively, and $x=E_{\gamma}/E_{e^-}$. The quasi-monochromatic photon
beam covered the energy range from 450 to 1450 MeV. The circular polarization depends
on the photon energy (Eq.~(\ref{eq:olsen})) and increased from $\approx$50\% at 450 MeV
to $\approx$80\% at 1450 MeV.

\begin{figure}
\begin{center}
\resizebox{0.48\textwidth}{!}{%
\includegraphics{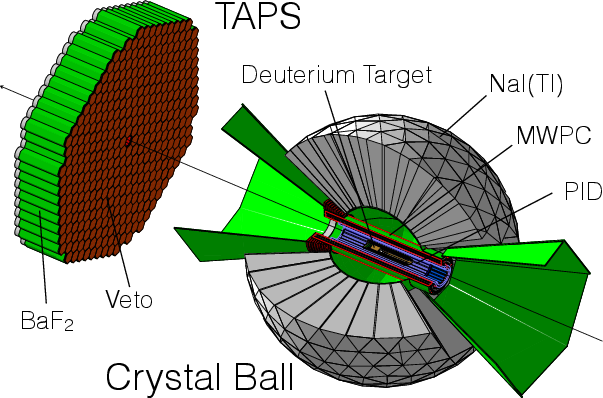}}
\caption{(Color online) Setup of the electromagnetic calorimeter combining the
Crystal Ball and TAPS detectors. Only three quarters of the CB are shown.
For charged particle identification mounted inside the CB were the Particle
Identification Detector (PID) and multiple wire chambers (MWPC) and in front
of TAPS the TAPS Charged-Particle Veto (CPV) detector. The beam enters from the bottom right
corner of the figure, the target was placed in the center of the CB}
\label{fig3}
\end{center}
\end{figure}

The experiment required transversely polarized protons, which were provided by a
frozen-spin butanol (${\rm C}_4{\rm H}_9{\rm OH}$) target. A specially designed
$^3$He/$^4$He dilution refrigerator was built for polarization measurements with
the CB detector. The target was cooled down to $\approx$20~mK in a strong magnetic
field of $\approx$2.5~T. Dynamic Nuclear Polarization (DNP) was used to transfer
the electron polarization to the free protons, bound in the butanol molecules,
by use of radio-frequency fields. When the required polarization was reached,
the magnet with the large field was removed and replaced by a small superconducting
holding coil (0.45~T), which allowed the target to be moved into the center of the main
detector. The holding coil was a four layer saddle coil which operated at a current 
of 35~A and held the transverse polarization with a relaxation time of around 1500~h. 
The low material budget of this coil guaranteed that the effects on the critical detection 
of charged particles (recoil protons) would be small. These effects were modeled 
using a Monte Carlo (MC) simulation. The cylindrical target container (2~cm long $\times$
2~cm diameter) was filled by 2~mm diameter butanol spheres with a filling factor around $60\%$. 
Typical polarization degrees obtained (averaged over the exponential decay due to the 
relaxation time) during the present measurements were around 70\%. More details about 
construction and operation of the target are given in Ref.\,\cite{Thomas_11}.

The use of butanol targets has the disadvantage that signals from unpolarized nucleons
bound in the carbon and oxygen nuclei of the molecule dilute the polarization signal.
The elimination of such backgrounds (as discussed in detail in the next section) requires
the additional measurement of the reaction from the free proton using liquid hydrogen
(LH$_2$) targets and from solid carbon targets. For the latter, a special solid carbon
foam was used that can be produced so that it matches exactly the geometry and the density
of carbon nuclei in the butanol target (in fact, the density was also corrected to
represent the much less abundant oxygen nuclei in butanol). In total, four different beam
times were analyzed for the present results, which are summarized in Tab.~\ref{tab:targets}.

\begin{table}[thh]
\begin{center}
\begin{tabular}{|c|c|c|c|c|}
\hline
& Run I & Run II & Run III & Run IV\\
\hline\hline
  Target type & LH$_2$ & C$_4$H$_9$OH & C$_4$H$_9$OH & C-foam\\
  target length [cm] & 10 & 2 & 2 & 2\\
  target radius [cm] & 4 & 2 & 2 & 2\\
  density [barn$^{-1}$] & 0.421 & 0.092 & 0.092 & 0.057\\
  e$^-$ energy [MeV] & 1558 & 1558 & 1558 & 1558\\
  beam-radius [cm] & 1.3 & 1.3 & 1.3 & 1.3\\
  multiplicity trigger & M3+ & M2+ & M2+ & M2+\\
  CB sum trigger [MeV] & 360 & 350 & 300 & 350\\
  MWPC & no & no & yes & yes\\
  PID & yes & yes & no & yes\\
\hline
\end{tabular}
  \caption[Summary of data sets]{
    \label{tab:targets}
     Main parameters of the experimental sets.
     The surface density in barn$^{-1}$ is given in nuclei/barn for the LH$_{2}$ target
     and the carbon-foam target, but in molecules/barn for the butanol targets. The effective
     surface densities of free protons in the butanol targets are a factor of 10 higher.
     The beam radius refers to the radius of the photon-beam spot size on the target.
}
\end{center}
\end{table}

Two production beam times with the polarized butanol target and one with a liquid
hydrogen and one with a carbon target for background elimination were taken. The trigger for
all measurements was based on a multiplicity condition for hits in the calorimeter.
For this purpose both CB and TAPS were subdivided into logical sectors
(see \cite{Prakhov_09,McNicoll_10,Werthmueller_14,Witthauer_17,Dieterle_18} for details).
However, this condition was not critical for the present analysis. Only the measurement
with the liquid hydrogen target used a multiplicity-three trigger, the other used
multiplicity-two triggers and a final state that requires four detected photons is not
biased by any of these conditions. In addition, an analog-sum threshold for the total energy
deposition in the CB between 300 and 360~MeV was required.

There are, however, two factors which complicated the analysis. The first is the difference
in target length between the measurement with the liquid hydrogen target and the measurements
with the solid targets. The resolution for invariant and missing masses (see next section),
used to separate signal from background, depends significantly on target length (due to the
uncertainty in the unmeasured reaction vertex). Results from the hydrogen measurement could
therefore not be directly compared to the solid state targets. Furthermore, the particle
identification detectors inside the CB, the PID and the MWPCs, were not fully operational during
all measurements. The MWPCs were not active for the measurement with the liquid hydrogen
target and for one of the beam-time periods with the butanol target (II), while the PID had
several dead channels for the other butanol measurement (III). This required some very
detailed data analysis which is discussed in the next section.


\section{Data analysis}\label{Data}

The data were analyzed in several different ways. All analyses accepted only events with
five hits (candidates for four photons and one proton) in the combined calorimeter.
The differences concerned the identification of the recoil protons and the treatment of
the carbon/oxygen background in the asymmetry ratios.

In all analyses photons and protons in TAPS were identified with the standard analysis methods
for this detector using the response of the TAPS CPV detector, a pulse shape analysis (PSA),
and a time-of-flight (ToF) versus-energy analysis as described in detail in
Refs.\,\cite{Dieterle_15,Werthmueller_14,Witthauer_17,Dieterle_18}.

Charged particles hitting the CB can be identified in principle from the response
of the PID and/or the MWPCs. There was, however, the problem that the MWPCs were not activated
for Runs I and II and the PID could not be used for Run III. Therefore the following analysis 
strategies were adopted.

All four runs were first analyzed ignoring the information from the PID and the MWPCs and
instead accepting only hits with one responding NaI crystal (hit multiplicity one)
in the CB as protons. In the energy range of interest, photon hits almost always activate two
or more modules. This analysis has the advantage that it minimizes instrumental asymmetries,
but it significantly reduces counting statistics because only a fraction of proton hits are
multiplicity-one hits. Furthermore, it suffers from low resolution for the proton angles, which
can be determined much more precisely when the MWPCs are used. All results from this analysis were 
averaged for the two butanol beam times (Runs II, III in Tab.~\ref{tab:targets}).

The other analyses used either the PID or the MWPCs for proton identification in the CB.
The PID was used for Run I and Run II, the MWPCs for Run III, and the carbon-background
measurement was in one case analyzed with the PID, ignoring the MWPC information and vice versa 
in the other case.  However, it turned out that for the analysis of Run II (first butanol beam time)
the efficiency calibration of the PID was not good enough to remove all instrumental effects
for the azimuthal distribution of the recoil protons. This was no problem for the extraction
of the double-polarization asymmetry $P_x^{\odot}$, because such artefacts cancel in
the difference of positive and negative beam polarization. However, the target asymmetry
$P_y$ was affected by it. Therefore, the analyses using the charged particle identification
detectors could be averaged over both butanol beam times for the $P_x^{\odot}$ observable, but
only Run III with the MWPCs was used for the target asymmetry $P_y$.

The other difference between the analyses was the treatment of the unpolarized background
from the carbon and oxygen nuclei in the butanol target. The asymmetries have been evaluated
from:
\begin{equation}
P_y{\rm sin}(\phi_s) =
\frac{1}{P_T}\frac{d\sigma^{\uparrow}(\phi_s)-d\sigma^{\downarrow}(\phi_s)}
{d\sigma^{\uparrow}(\phi_s)+d\sigma^{\downarrow}(\phi_s)}
\label{eq:T}
\end{equation}
and from:
\begin{equation}
P_x^{\odot}{\rm cos}(\phi_s) =
\frac{1}{P_T}\frac{1}{P_{\odot}}
\frac{d\sigma^{+}(\phi_s)-d\sigma^{-}(\phi_s)}{d\sigma^{+}(\phi_s)+d\sigma^{-}(\phi_s)}\;,
\label{eq:F}
\end{equation}
with
\begin{eqnarray}
d\sigma^{\uparrow} & \equiv  d\sigma^{\uparrow +} + d\sigma^{\uparrow -},\;\;\;\; &
d\sigma^{\downarrow} \equiv  d\sigma^{\downarrow +} + d\sigma^{\downarrow -}\\
d\sigma^{+} & \equiv  d\sigma^{\uparrow +} + d\sigma^{\downarrow -},\;\;\;\; &
d\sigma^{-} \equiv  d\sigma^{\downarrow +} + d\sigma^{\uparrow -}\nonumber
\end{eqnarray}
where $P_T$ and $P_{\odot}$ are the polarization of the target (transverse) and
beam (circular), respectively. The cross sections refer to the different orientations
of the target spin ($\uparrow$, $\downarrow$) and the two polarization states of the beam
($+$, $-$).

Backgrounds from the non-polarized nucleons bound in the carbon and oxygen nuclei of the
butanol targets cancel in Eqs.~\ref{eq:T},\ref{eq:F} in the numerator, but they contribute
in the denominator. This background contribution can be treated in two different ways.
Either, the contribution from the nucleons bound in carbon/oxygen nuclei is removed with
the help of the background measurement using a carbon foam target (the small contribution
from oxygen was approximated by an $A^{2/3}$ scaling law \cite{Maghrbi_13}). Alternatively,
the results from the measurement with a liquid hydrogen target, which are free of such
backgrounds, can be used in the denominator. In the present work both methods were exploited
for the analyses using the charged particle identification detectors, but only the carbon
subtraction method was used for the analysis relying on proton identification via hit
multiplicity.

\begin{table}[hth]
\begin{center}
\begin{tabular}{|c|c|c|c|}
\hline
Analyses & (1) & (2) & (3)\\
\hline\hline
 proton & MWPC/(PID) & MWPC/(PID) & hit multiplicity\\
 denominator & carbon sub. & hydrogen norm. & carbon sub.\\

\hline
\end{tabular}
  \caption[Summary of data sets]{
    \label{tab:ana}
    Characteristics of the analyses (1), (2), (3). First line: identification of recoil
    proton by charged particle detectors or hit multiplicity in CB.
    Second line: treatment of non-polarized background, carbon subtraction or denominator of
    asymmetry from measurement with liquid hydrogen target.
}
\end{center}
\end{table}

The analysis strategies are summarized in Tab.~\ref{tab:ana}.
The advantage of the analyses using the carbon subtraction is that many systematic
uncertainties cancel in the ratios of Eqs.~\ref{eq:T},\ref{eq:F}. The advantage of the
normalization to the results from a hydrogen target in Eqs.~\ref{eq:T},\ref{eq:F}
is the elimination of unpolarized backgrounds.

The main steps for the identification of decay photons and recoil protons from the
$\gamma p\rightarrow p\pi^0\pi^0$ reaction were similar to the methods used in
\cite{Oberle_13,Dieterle_15,Kashevarov_09}. In the first step, hits in the calorimeter
were designated as `charged' or `neutral' using either the information from the TAPS CPV
and the PID and/or the MWPCs surrounding the target (analyses (1),(2)) or the hit
multiplicity (analysis (3)).

In the next step, the four photon candidates were analyzed. The relative timing between
the photons was measured for photon pairs in TAPS with a resolution (FWHM) of $\approx$0.5~ns,
for photon pairs with one photon in TAPS and one in CB with $\approx$1.5~ns, and for pairs
in CB with $\approx$2.5~ns. Cuts were applied to the relative timing, however they had
very little effect. The initial spectra already had a very low background level
and were background free after the subsequent kinematic cuts were applied.
The timing of the photons was also used to remove background from random tagger-calorimeter
coincidences. In this case the resolution was $\approx$1~ns for the TAPS-tagger coincidence
and $\approx$1.5~ns for the CB-tagger coincidence. The random background was removed as in
previous analyses (see e.g. \cite{Werthmueller_14}) by a cut on the prompt time peak and a
side-band subtraction of the flat background.

A $\chi^2$ analysis was used to identify the most probable out of the three combinatorial
possible combinations of the four photons from the decay of two $\pi^0$ mesons. The $\chi^2$ 
was defined by
\begin{equation}
\label{eq:chi2}
\chi^{2}(k) = \sum_{i=1}^{2}\left
(\frac{m_{\pi^0}-m_{i,k}}{\Delta m_{i,k}}\right)^{2} ~~{\rm with}~~ k=1,..,3\;,
\end{equation}
where $m_{\pi^0}$ is the nominal $\pi^0$ mass, the $m_{i,k}$ are the invariant masses of
the $i$-th pair in the $k$-th permutation of the hits and $\Delta m_{i,k}$ is the corresponding
uncertainty from the experimental energy and angular resolution. Both were computed
event-by-event. Only the combination with the smallest $\chi^2$ was analyzed further.

\begin{figure*}
\begin{center}
\resizebox{1.0\textwidth}{!}{%
\includegraphics{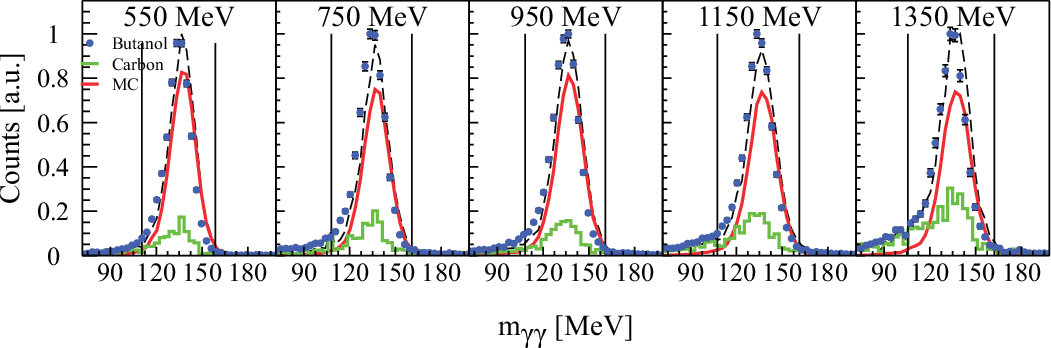}
\hspace*{0.5cm}\includegraphics{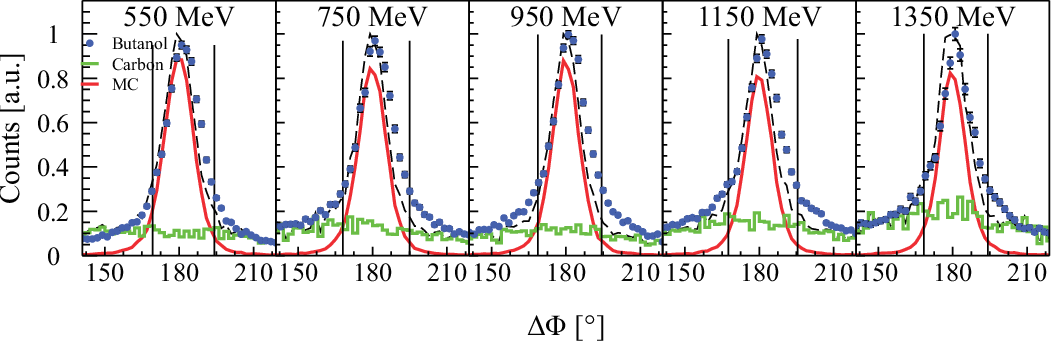}}
\caption{(Color online) Left hand side: invariant masses for different ranges of incident photon
energy (indicated in figure). Points (blue) with error bars: measured butanol data.
Solid (red) lines: hydrogen data corrected by MC simulation for resolution effects (see text).
Solid (green) lines: background from carbon data.
Dashed (black) lines: sum of hydrogen and carbon data. Vertical lines indicate cuts applied
to invariant mass. The cuts on coplanarity and missing mass are applied to this spectra.
Right hand side: coplanarity spectra (see text). Same notation as left hand side.
Cuts on invariant mass are applied. All butanol data from Run III, analyzed with charged particle 
detectors (MWPCs). Results from Run II (PID) are not significantly different for the invariant mass,
but have less background in coplanarity, in particular outside the peak region.}
\label{fig:invmass}
\end{center}
\end{figure*}

\begin{figure*}
\begin{center}
\resizebox{1.0\textwidth}{!}{%
\includegraphics{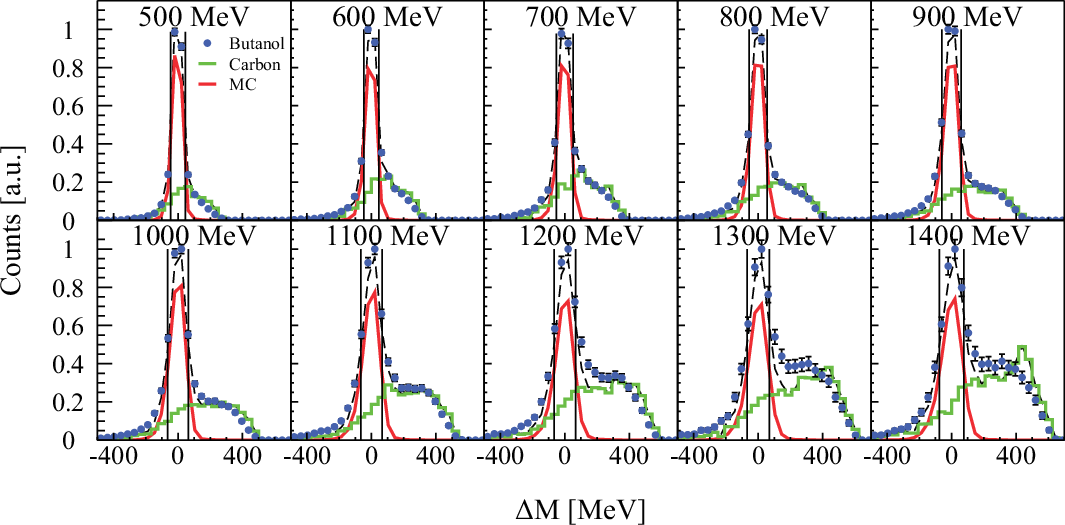}
\hspace*{0.5cm}\includegraphics{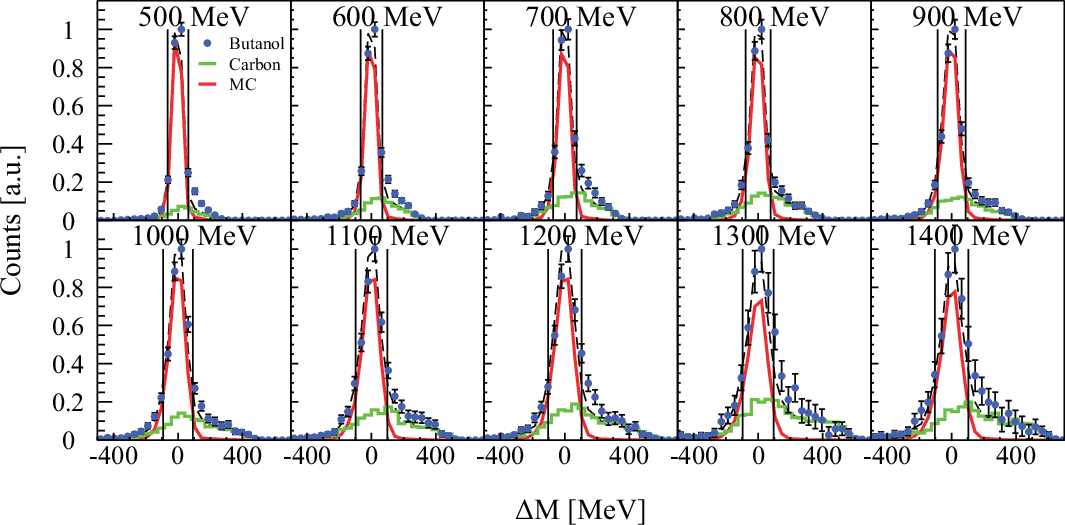}}
\caption{(Color online) Missing mass spectra for different ranges of incident photon energy.
Notation as in Fig.~\ref{fig:invmass}. Left hand side: Run III (charged particle identification
with MWPCs), right hand side: Run II (charged particle identification with PID).
Vertical lines indicate cuts applied to missing mass. The invariant mass cuts are applied to
this spectra.
}
\label{fig:mismas}
\end{center}
\end{figure*}

\begin{figure}
\begin{center}
\resizebox{0.48\textwidth}{!}{%
\includegraphics{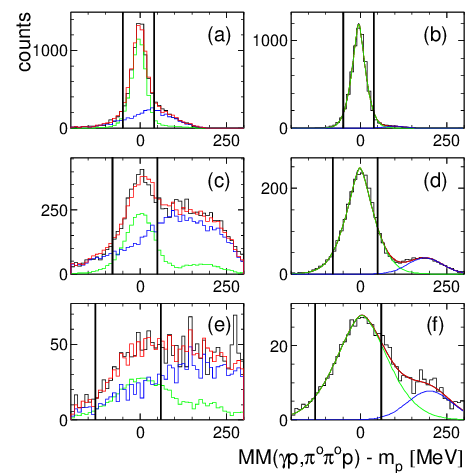}}
\caption{(Color online) Left hand side ((a),(c),(e): typical missing mass spectra from
analysis (3) with proton identification by hit multiplicity for photon energies around
600, 900, and 1400~MeV (bin widths $\pm$50~MeV). Black histograms: butanol data,
green histograms: hydrogen data, blue histograms: carbon data, red histograms: sum of
hydrogen and carbon data. Right hand side ((b),(d),(f))): missing mass spectra for
hydrogen target. Black histograms: measured data, green lines: MC simulation
of $\pi^0\pi^0$ final state, blue lines: MC simulation of background from $3\pi^0$
final states, red lines: sum of signal and background from MC.}
\label{fig:vicana}
\end{center}
\end{figure}

In the next step the invariant masses of these photon pairs were analyzed. Figure~\ref{fig:invmass}
shows some typical examples of the invariant mass of the photon pairs obtained from Run III
with the butanol target compared to the results measured with the carbon foam and hydrogen
target. The analysis was done with the use of the MWPCs for identification of the proton
(results from Run II with usage of the PID are very similar). The counts from the three different
targets (butanol, carbon, hydrogen) have been absolutely normalized by the target densities, integrated
photon fluxes, and the detection efficiencies obtained from Monte Carlo (MC) simulations done with the
Geant4 code \cite{Geant4}. The simulations included known inefficiencies of the charged particle
detectors. The results from the measurement with the hydrogen target could not be
compared directly to the other two targets because the length of the hydrogen target
(see Tab.~\ref{tab:targets}) was five times the length of the butanol and carbon targets which
affects the experimental resolution because the reaction vertex is not known. Therefore, in a first
step the results from the hydrogen target were compared to a MC simulation using the actual length of
that target. Almost perfect agreement of the line shape was found. Subsequently, the MC simulation for
the hydrogen target was repeated for the target length of the butanol target and the simulated line
shape was normalized with the measured count rate of the hydrogen target. This was also done for the
kinematical spectra (coplanarity and missing mass) discussed below. After this correction the
invariant-mass line shapes for the three targets were very similar as expected because nuclear
Fermi motion does not affect invariant masses. The figure shows only a few examples of these
spectra for different energy bins. The actual analysis was carried out using finer bins of energy and
also as a function of the polar angle of the recoil proton. The sum of the absolutely normalized
count rates for the hydrogen and carbon target reproduces very well the measured yield for the
butanol target. The vertical lines in the figure indicate the cuts applied in further analysis.

After the invariant-mass analysis, the mass of the $\pi^0$ meson was used as a constraint
to improve the resolution in further kinematical analyses. Since the angular resolution of
the detector is much better than the energy resolution this was simply done by replacing the
measured energies of the decay photons $E_{\gamma_1}$, $E_{\gamma_2}$ by
\begin{equation}
E'_{\gamma_1,\gamma_2} = \frac{m_{\pi^0}}{m_{\gamma_1\gamma_2}}E_{\gamma_1,\gamma_2}\;,
\end{equation}
where $m_{\pi^0}$ is the mass of the $\pi^0$ meson and $m_{\gamma_1\gamma_2}$ is the invariant
mass of the photon pair with originally measured energies.

The first kinematic condition that was checked is the coplanarity of the two-pion system and
the recoil proton. Due to momentum conservation, the difference in the azimuthal angle
$\Delta\Phi$ between the $\pi^0\pi^0$ pair and the recoil nucleon must be 180$^{\circ}$.
This is normally not the case when additional particles have escaped detection (for example from
triple-pion final states). The result of this analysis for Run III (Run II is very similar)
is shown in Fig.~\ref{fig:invmass} (right hand side) together with the applied cuts. Again the
normalized results from the hydrogen and carbon target add up to the data obtained with the
butanol target. In this case the carbon background is rather flat due to the large effect from
nuclear Fermi motion.

Even more efficient is the analysis of the missing mass for which the recoil proton, although
detected, is treated as a missing particle and its four momentum is reconstructed from the reaction
kinematic using
\begin{equation}
\Delta M = \left|P_{\gamma}+P_{N}-P_{\pi^0_1}-P_{\pi^0_2}\right| -m_N\ ,
\label{eq:mm}
\end{equation}
where  $P_{\gamma}$, $P_{N}$, $P_{\pi^0_1}$, $P_{\pi^0_2}$ are the four-momenta of the incident
photon, the initial state nucleon (at rest) and the two pions. The nucleon mass $m_{N}$ was
subtracted so that true $\gamma N\rightarrow N\pi^0\pi^0$ events were expected at $\Delta M = 0$.
The result of this analysis is shown in Fig.~\ref{fig:mismas} for the analyses of Run III (left
hand side) and Run II (right hand side). In both cases the normalized yields from the hydrogen and
carbon targets add up to the butanol data. The background suppression is better for the measurement
using the PID detector (Run II) than with the MWPCs (Run III). However, as mentioned earlier, the
response for the proton detection was more isotropic in the azimuthal angle $\Phi$ for the
measurement with the MWPCs. These spectra (again analyzed with finer energy bins as a function
of the proton polar angle) were used to subtract the carbon background in analysis (1) (see
table~\ref{tab:ana}).

In the case of analysis (2) the carbon background played no role because it cancels in the numerator 
of the asymmetry and the measurement with the hydrogen target was directly used for normalization.

Missing mass spectra for analysis (3), using the hit multiplicity for proton identification, are
summarized in Fig.~\ref{fig:vicana}. The left-hand side of the figure shows the comparison of
the yields from the butanol, carbon, and hydrogen targets. In this case the three yields were not
absolutely normalized but the relative contribution of the carbon and hydrogen target was fitted
so that the sum reproduced the butanol measurement. The fitting was only done in the signal
missing-mass range indicated by the vertical lines. The right-hand side of the figure shows the
experimental results for the hydrogen target compared to MC simulated responses for double $\pi^0$
production and background from 3$\pi^0$ final states. Also in this analysis background from carbon
was subtracted and the background from $3\pi^0$ production does not significantly
intrude into the missing mass range selected by the cuts. Some results for the two asymmetries 
$P_x$ and $P_y^{\odot}$, which were not investigated with analysis (1) and (2), were obtained from 
this analysis.

\section{Systematic uncertainties}\label{sec:systematics}

Systematic uncertainties arise from several sources with different impact on the analysis
strategies and the investigated observables. The measurement of the target polarization degree
$P_T$ affects all measured asymmetries in the same way. The polarization of the target was
regularly reversed in order to reduce the uncertainty. The systematic uncertainty in the target
polarization was estimated from the NMR measurements to be $\approx$4\%. The polarization degree 
of the photon beam affects only the measurement of the $P_x^{\odot}$ and $P_y^{\odot}$ beam-target 
asymmetries. This was estimated from the Mott measurements to be $\approx$3\%. The measurement of 
the various asymmetries is also affected by variations in the azimuthal efficiency of the detectors.
The detector response was MC simulated and corrected. Residual effects were investigated in two ways. 
\begin{figure}[!thb]
\begin{center}
\resizebox{0.49\textwidth}{!}{%
\includegraphics{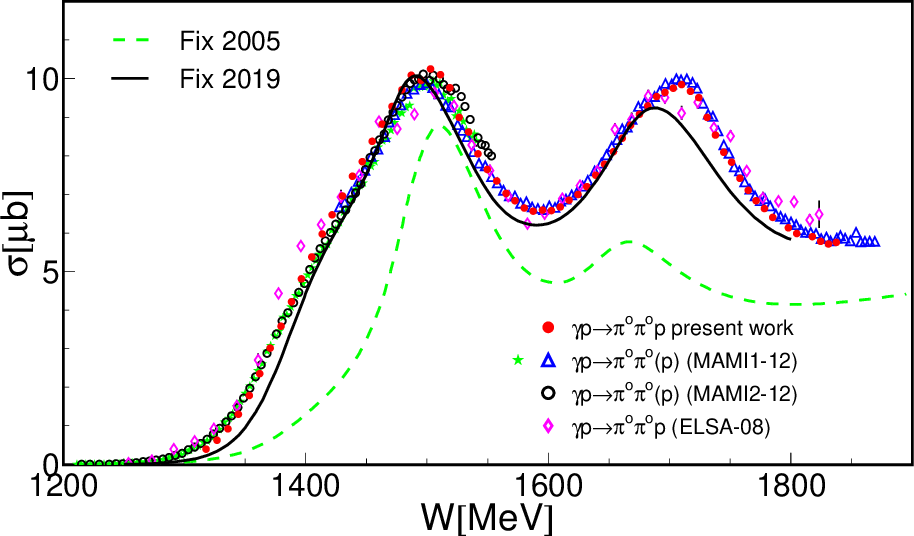}}
\caption{(Color online) Total cross section for the $\gamma p\rightarrow p\pi^0\pi^0$
reaction as function of the total cm energy $W$. Data from the present analysis (red, filled spheres)
for the measurement with a liquid hydrogen target compared to previous results from the Mainz
MAMI facility (blue, open triangles and green stars \cite{Kashevarov_12}; black open circles
\cite{Zehr_12}) and the Bonn ELSA facility \cite{Thoma_08}). The (green) dashed line is the
prediction for the total cross section from the MAID model \cite{Fix_05}. The (black) solid line
is an update of the same model using nucleon resonance parameters from the 2018 PDG compilation
\cite{PDG_18} and a refit of the model to all available data.}
\label{fig:tot_hyd}
\end{center}
\end{figure}
The target asymmetry was determined individually for both directions of target polarization by 
an analysis of the azimuthal angular distributions replacing $d\sigma(\phi_s)^{\downarrow}$
by $d\sigma(\phi_s+\pi)^{\uparrow}$ in Eq.~(\ref{eq:T}) etc. In the absence of instrumental effects 
both results must agree. Furthermore, the target asymmetries were fitted with an incorrect 
cos$(\phi_s)$ angular dependence in which case they must vanish unless instrumental asymmetries 
contribute. Both conditions were fulfilled within statistical uncertainties for Run III using the 
MWPCs, but not for Run II using the PID. Therefore, Run II, which was also inferior in terms of 
counting statistics, was discarded for the target asymmetries. Such effects are not significant 
for the beam-target asymmetries. They cancel in the subtraction of the count rates from the 
two beam polarizations.

Further systematic uncertainty arises from either the subtraction of the carbon background in 
analyses (1,3) or the normalization to hydrogen data in analysis (2). When the denominator of 
Eqs.~\ref{eq:T},\ref{eq:F} is taken from the measurement with a hydrogen target all cross sections 
must be absolutely normalized (beam flux, target density, detection efficiency). For the subtraction 
of the carbon contribution the relative normalization of the butanol and carbon data matters. 
The target size and density of the butanol and carbon targets were well matched (the density of the 
carbon foam was identical to the density of carbon nuclei in the butanol target including a correction 
for oxygen nuclei). Also the experimental conditions (trigger, thresholds) were kept as similar as
possible, so that many effects canceled without requiring a precise normalization. The largest
effects were then due to the exact reproduction of line shapes in the missing-mass spectra used
to eliminate the background. This depends on the methods used to identify the recoil proton
(PID, MWPCs, or hit multiplicity, see Figs.~\ref{fig:mismas},\ref{fig:vicana}).

\begin{figure}[!thb]
\begin{center}
\resizebox{0.49\textwidth}{!}{%
\includegraphics{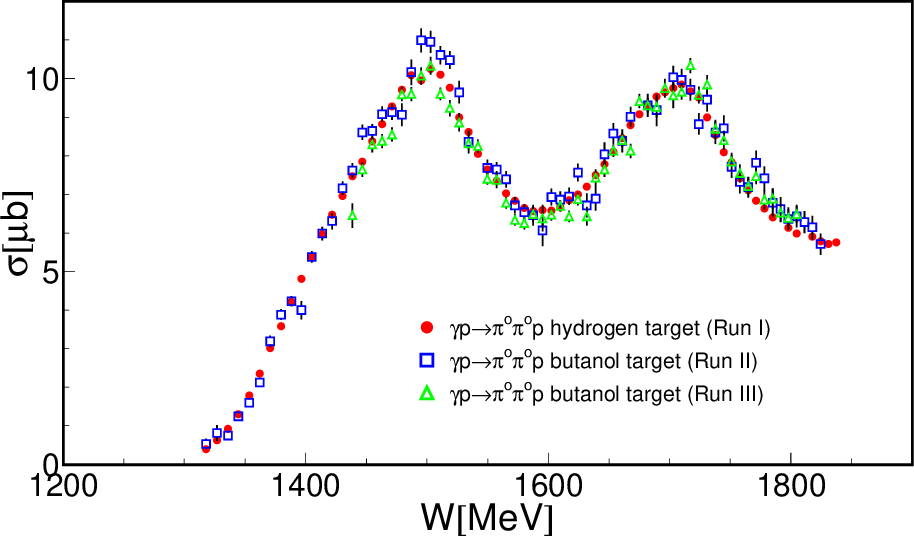}}
\caption{(Color online) Comparison of the total cross section for the hydrogen target (Run I)
from the present analysis to the total cross sections for the $\gamma p\rightarrow p\pi^0\pi^0$
extracted from the measurements with the butanol targets (Runs II,III) after subtraction of
the carbon background from Run IV.}
\label{fig:tot_but}
\end{center}
\end{figure}

The absolute normalization of the data included not only target density and photon
fluxes. Also the instrumental acceptance and detection efficiency were tested by an analysis
of the total cross section of the $\gamma p\rightarrow p\pi^0\pi^0$ reaction.
Fig.~\ref{fig:tot_hyd} shows the total cross section of this reaction from the present
measurement with the hydrogen target (Run I) compared to previous results. The statistical quality
of the present data is even better than that of the previous measurements. No significant systematic
deviations were observed. This is not trivial because for the previous MAMI results
\cite{Kashevarov_12,Zehr_12} only detection of the four $\pi^0$ decay photons, with subsequent
invariant-mass and missing-mass analyses, was required. In contrast to the present analysis, recoil
protons were ignored in these measurements. The previous analysis strategy avoided all problems
with proton detection efficiency. However, for the measurements with the butanol target coincident
detection of recoil protons is very useful for background elimination. Therefore, the
hydrogen data were analyzed in coincidence with protons using the charged particle detectors (PID, CPV)
Tof-versus-energy, and PSA.

The curves in Fig.~\ref{fig:tot_hyd} show the predictions of the isobar model from Ref.~\cite{Fix_05}, 
which was used as basis of the analysis of the data. The dashed line shows
the result from Ref.~\cite{Fix_05}, the solid line the update with resonance parameters from the
most recent PDG (Particle Data Group) \cite{PDG_18} compilation and a refit to double pion 
production data. This refit describes the data much better than the original version of
the model.

Figure~\ref{fig:tot_but} shows a comparison of the total cross section from the present analysis
of the measurement with the hydrogen target (Run I) to the results from the two measurements
(Run II, III) with the butanol target after subtraction of the carbon background determined with
Run~IV. For these analyses proton identification was done with the charged particle detectors active
for runs I,II,III (see Tab.~\ref{tab:targets}).  

%
%
%
%
\begin{figure*}[!thb]
\begin{center}
\resizebox{0.85\textwidth}{!}{%
\includegraphics{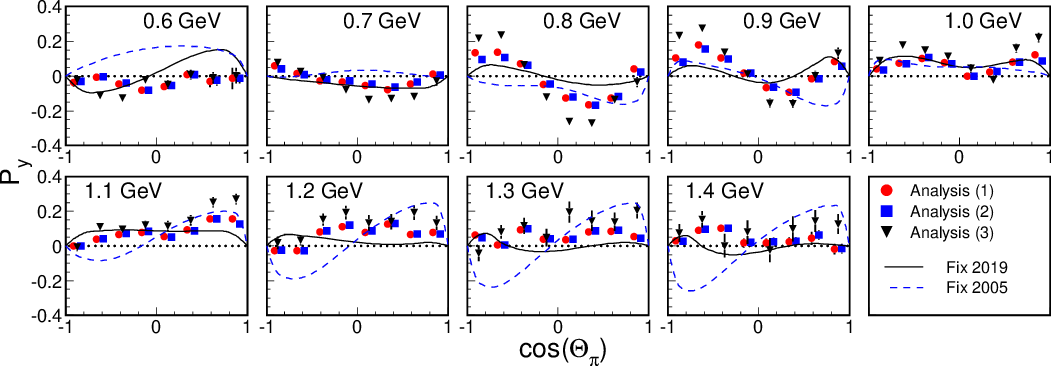}}
\resizebox{0.85\textwidth}{!}{%
\includegraphics{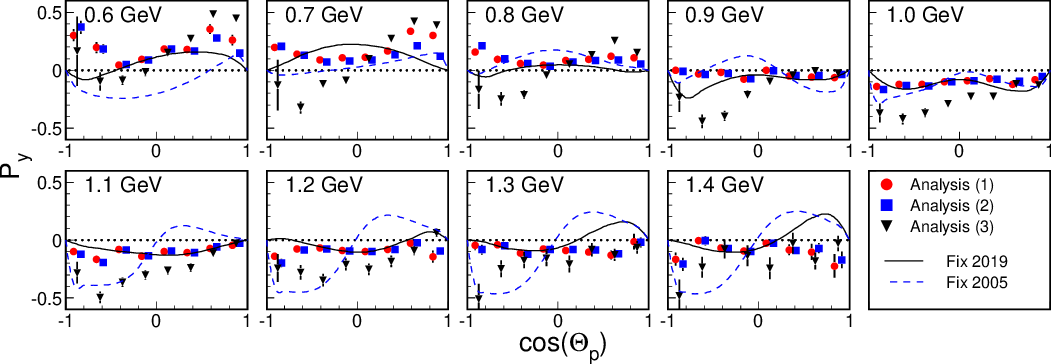}}
\resizebox{0.85\textwidth}{!}{%
\includegraphics{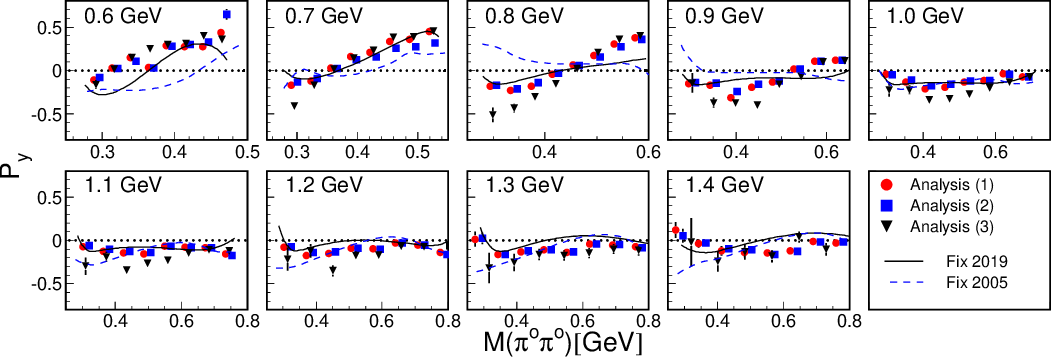}}
\resizebox{0.85\textwidth}{!}{%
\includegraphics{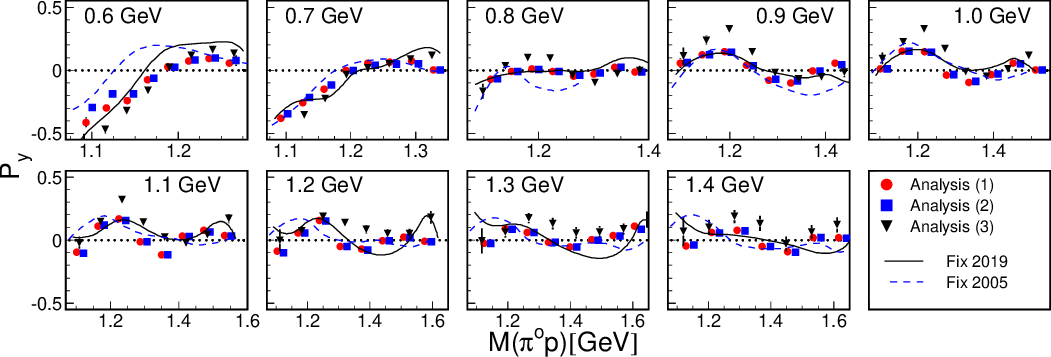}}
\caption{(Color online)
Angular and invariant mass distributions for the target asymmetry
$P_y$ of the $\gamma p\to\pi^0\pi^0 p$ cross section for incident photon energies from
600 to 1400 MeV (bin centroids). From top to bottom results as function of pion polar angle
(both pions for each event), proton polar angle, invariant mass of pion pair and invariant
mass of nucleon pion pairs (two entries per event). The (red) points correspond to analysis (1),
the (blue) squares to analysis (2), and the (black) triangles to analysis (3).
The curves show the predictions of the isobar model \protect\cite{Fix_05}.
The (blue) dashed curves correspond to the original 2005 version, the (black) solid curves to
a refit including the new results (see text for more details).}
\label{fig:Tasym}
\end{center}
\end{figure*}
%
%
%
%
\begin{figure*}[!htb]
\begin{center}
\resizebox{0.85\textwidth}{!}{%
\includegraphics{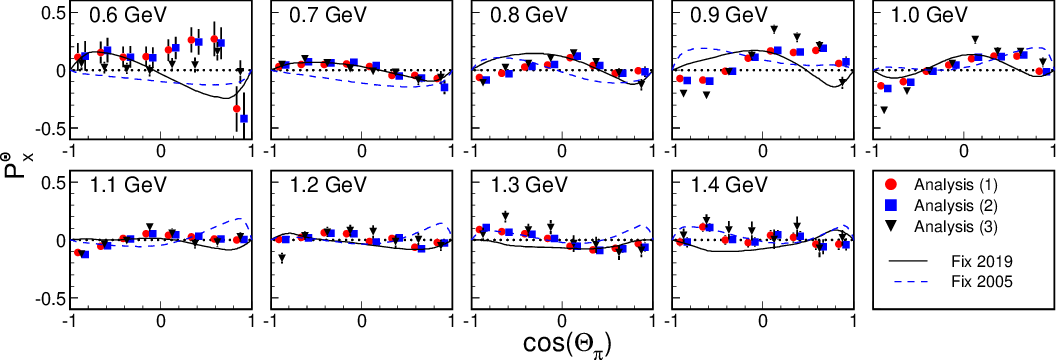}}
\resizebox{0.85\textwidth}{!}{%
\includegraphics{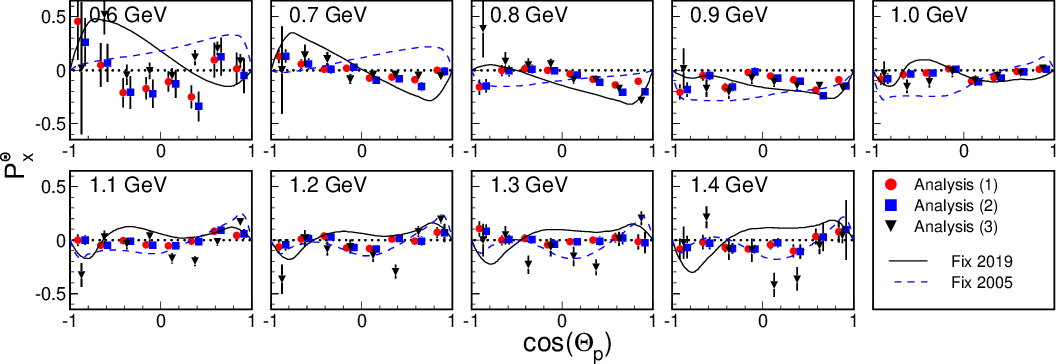}}
\resizebox{0.85\textwidth}{!}{%
\includegraphics{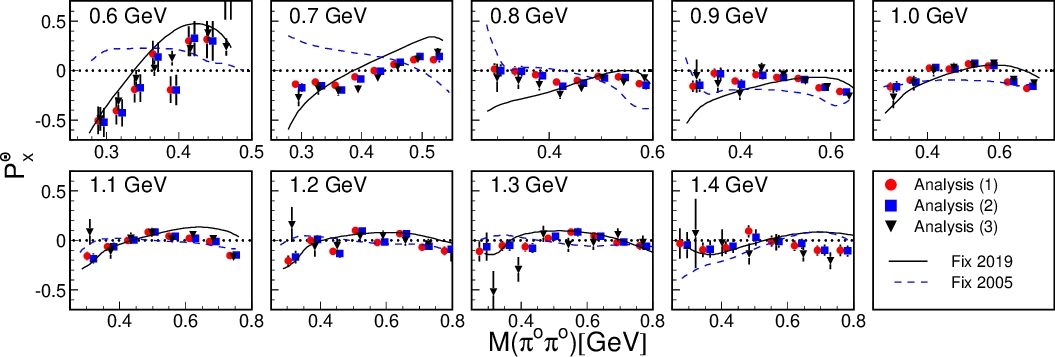}}
\resizebox{0.85\textwidth}{!}{%
\includegraphics{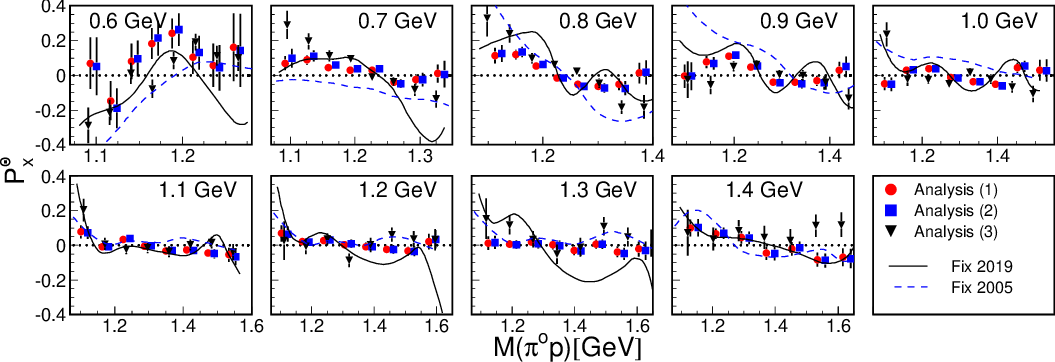}}
\caption{(Color online)
Same as in Fig.\,\protect\ref{fig:Tasym} for the beam-target asymmetry $P_x^\odot$.}
\label{fig:Fasym}
\end{center}
\end{figure*}
\clearpage

The carbon data was analyzed with the PID when it was compared to Run~II and with the MWPCs for 
comparison with Run~III. The agreement is reasonably good and demonstrates that the absolute 
calibration and the background subtraction procedure allows the correct cross section for the free 
$\gamma p\rightarrow p\pi^0\pi^0$ reaction to be extracted from the measurements with the solid 
butanol targets. Therefore, systematic effects on the asymmetries from the normalization of the
cross section data and the background subtraction are not expected to be important. This can be
verified from the comparison of the asymmetries extracted with the hydrogen normalization method
or the carbon background subtraction (see Sec.~\ref{sec:discussion}).

\section{Discussion of the results}\label{sec:discussion}

\begin{figure*}[!thb]
\begin{center}
\resizebox{0.94\textwidth}{!}{%
\includegraphics{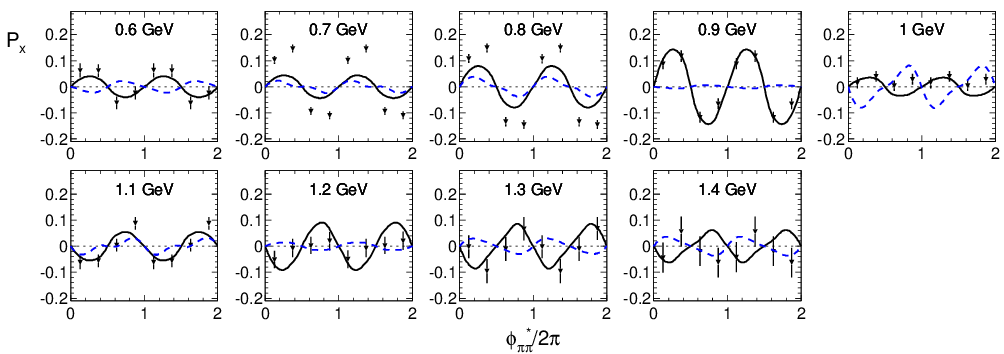}}
\caption{(Color online) Same as in Fig.\,\protect\ref{fig:Tasym} for the asymmetries
$P_x(\phi^*_{\pi\pi})$ from analysis (3). ($\phi^*_{\pi\pi}$ angle between production and reaction plane,
see Fig.~\ref{fig:frame}).}
\label{fig:PX}
\end{center}
\end{figure*}

\begin{figure*}[!thb]
\begin{center}
\resizebox{0.94\textwidth}{!}{%
\includegraphics{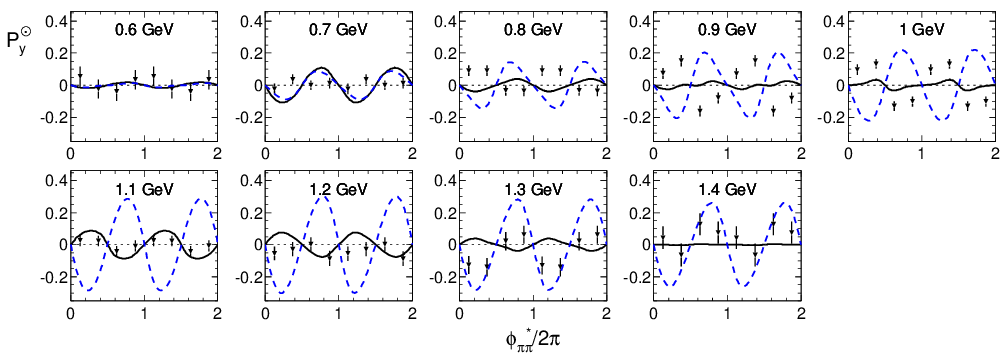}}
\caption{(Color online) Same as in Fig.\,\protect\ref{fig:Tasym} for the asymmetry
$P_y^\odot(\phi^*_{\pi\pi})$ from analysis (3). ($\phi^*_{\pi\pi}$ angle between production and reaction plane,
see Fig.~\ref{fig:frame}).}
\label{fig:PYC}
\end{center}
\end{figure*}

Figures~\ref{fig:Tasym} - \ref{fig:PYC} show the results for the asymmetries. We follow
the definitions of the asymmetries introduced in Ref.\,\cite{Roberts_05}. Those used in
Ref.\,\cite{Fix_11} may be obtained from Eqs.\,(\ref{eq2}). The target asymmetries $P_y$ and
the beam-target asymmetries $P_x^{\odot}$ extracted from the three different analyses are
compared in Figs.~\ref{fig:Tasym} and \ref{fig:Fasym}, respectively.
The results from the two analyses using the charged particle detectors and either a normalization
to the measurement with the liquid hydrogen target or the subtraction of the carbon background
are in quite good agreement.

The analysis with the crude proton identification by hit multiplicity agrees reasonably well 
with the other two results for the beam-target double asymmetry $P_x^{\odot}$ (Fig.~\ref{fig:Fasym})
for which many systematic effects cancel. Agreement for the target asymmetry $P_y$ is also 
acceptable for the asymmetries as function of invariant masses and of pion angles, but not
for the asymmetries as function of proton angle. In the latter case, the results not using the MWPCs
suffer not only from misidentifications but also from the much worse angular resolution
for the recoil protons. We have therefore discarded them in Fig.~\ref{fig:Tasym}.
Overall, the statistical precision of analysis (3) is much lower than for the other two
analyses because the cut on hit-multiplicity one for the proton candidates eliminates
a significant fraction of good events.

\begin{table*}
\renewcommand{\arraystretch}{1.2}
\caption{Parameters of the $N$-type resonances fitted to the data for $\gamma N\to\pi\pi N$ in the 
region up to the total center-of-mass energy $W=1900$ MeV. $M_R$ and $\Gamma_{tot}$ are the Breit-Wigner 
mass and total width of a resonance. The notations $\beta_\alpha$ are used for the branching ratios 
$\Gamma_\alpha/\Gamma_{tot}$ with $\alpha=\{\pi\Delta,\,\rho N,\,\sigma N\}$. The total widths 
$\Gamma_{tot}$ were not varied. Their values were taken directly from the Particle Data Group tables 
\cite{PDG_18}. In the second row for each resonance the parameters from the earlier analysis 
\protect\cite{Fix_05} are given. The third row lists the corresponding numbers from the compilation of 
\cite{PDG_18}. In those cases, when no estimation of $A^{(p)}_{1/2}$ and $A^{(p)}_{3/2}$ is made in 
\cite{PDG_18} their values were taken from the analyses of \cite{Sokhoyan_15b} and \cite{Anisovich_11} 
(marked with an asterisk).}
\label{ta2}
\begin{center}
\begin{tabular*}{17.4cm}
{@{\hspace{0.3cm}}c@{\hspace{0.3cm}}|@{\hspace{0.3cm}}c@{\hspace{0.3cm}}|
 @{\hspace{0.3cm}}c@{\hspace{0.3cm}}|@{\hspace{0.3cm}}c@{\hspace{0.3cm}}|
 @{\hspace{0.3cm}}c@{\hspace{0.3cm}}|@{\hspace{0.3cm}}c@{\hspace{0.3cm}}|
 @{\hspace{0.3cm}}c@{\hspace{0.3cm}}
}
\hline\hline\noalign{\smallskip}
 $N(M_R)J^P$ & $M_R$ & $\Gamma_{tot}$
& $\sqrt{\beta_{\pi\Delta}}\,A^{(p)}_{1/2}$ & $\sqrt{\beta_{\rho N}}\,A^{(p)}_{1/2}$ & $\sqrt{\beta_{\sigma N}}\,A^{(p)}_{1/2}$ & $A^{(p)}_{3/2}/A^{(p)}_{1/2}$ \\
  & [MeV] & [MeV] & [10$^{-3}$GeV$^{-1/2}$] & [10$^{-3}$GeV$^{-1/2}$] & [10$^{-3}$GeV$^{-1/2}$] & \\
\noalign{\smallskip}\hline\hline\noalign{\smallskip}
$N(1440)\frac12^+$  & 1449 & 350 & $-46.7$ &$-12.7$ & $-13.4$&  \\
                    & 1440 & 350 & $-30.4$ &$0$ & $-23.5$&  \\
                    & $1440\pm30$ & $350\pm100$ & $-26.4\pm16.0$ &$-$ & $-26.8\pm9.7$&  \\
                    \hline\noalign{\smallskip}
$N(1520)\frac32^-$  & 1525 & 110 & $-3.98$ &$-9.98$ & $-1.51$ & $-7.17$ \\
                    & 1515 & 115 & $-8.94$ &$-8.94$ & $0$ & $-7$ \\
                    & $1515\pm5$ & $110\pm10$ & $-11.9\pm1.5$ &$-$ & $-2.25\pm1.50$& $-6.22\pm1.85$ \\
                    \hline\noalign{\smallskip}
$N(1535)\frac12^-$  & 1538 & 150 & $42.8$ &$37.5$ & $32.1$ &  \\
                    & 1535 & 150 & $0$ & $17.78$ & $17.78$ & \\
                    & $1530\pm15$ & $150\pm25$ & $16.6\pm15.2$ &$-$ & $25.7\pm24.7$&  \\
                    \hline\noalign{\smallskip}
$N(1650)\frac12^-$  & 1635 & 125 & $18.4$ &$22.7$ & $7.85$ &  \\
                    & $-$  & $-$  & $-$  & $-$  & $-$  &   \\
                    & $1650\pm15$ & $125\pm25$ & $15.6\pm14.5$ &$-$ & $14.2\pm15.9$&  \\
                    \hline\noalign{\smallskip}
$N(1675)\frac52^-$  & 1660 & 145 & $10.8$ &$4.45$ & $8.32$ & $1.01$ \\
                    & 1675 & 150 & $14.7$ &$0$ & $0$& $1.05$ \\
                    & $1675^{+5}_{-10}$ & $145\pm15$ & $9.58\pm8.73$ &$-$ & $3.91\pm4.15$& $1.28\pm0.98$ \\
                    \hline\noalign{\smallskip}
$N(1680)\frac52^+$  & 1695 & 120 & $-1.77$ & $-2.15$ & $-1.30$ & $-17.0$ \\
                    & 1685 & 130 & $-4.68$ & $-4.93$ & $-5.17$& $-8.87$ \\
                    & $1685^\pm5$ & $120^{+10}_{-5}$ & $-4.74\pm0.14$ &$-$ & $-4.30\pm0.14$& $-11.7\pm6.2$ \\
                    \hline\noalign{\smallskip}
$N(1700)\frac32^-$  & 1732 & 200 & $31.6$ &$17.2$ & $6.49$ & $-0.75$ \\
                    & $-$ & $-$  & $-$  & $-$  & $-$  & $-$  \\
                    & $1720^{+80}_{-70}$ & $200\pm100$ & $34.3\pm30.1^*$  & $25.2\pm20.5^*$ & $11.6\pm14.8^*$ & $-0.90\pm0.03^*$\\
                    \hline\noalign{\smallskip}
$N(1710)\frac12^+$  & 1711 & 140 & $40.4$ &$11.2$ & $2.58$ &  \\
                    & $-$ & $-$  & $-$  & $-$  & $-$  &   \\
                    & $1710\pm30$ & $140\pm60$ & $12.2\pm11.1^*$  & $20.6\pm16.4^*$ & $-$ &  \\
                    \hline\noalign{\smallskip}
$N(1720)\frac32^+$  & 1723 & 250 & $77.7$ &$24.1$ & $5.65$ & $1.12$ \\
                    & $-$ & $-$  & $-$  & $-$  & $-$  & $-$  \\
                    & $1720^{+30}_{-40}$ & $250^{+150}_{-100}$ & $82.4\pm62.3$ &$12.2\pm9.5$ & $28.3\pm30.1$& $1.35\pm0.67^*$ \\
                    \hline\noalign{\smallskip}
$N(1900)\frac32^+$  & 1920 & 200 & $21.5$ &$0$ & $6.09$ & $-2.79$ \\
                    & $-$ & $-$  & $-$  & $-$  & $-$  & $-$  \\
                    & $1920\pm30$ & $200^{+120}_{-100}$  & $17.0\pm20.6^*$ & $-$ & $4.80\pm6.96^*$ & $-2.79\pm0.38^*$ \\
\noalign{\smallskip}\hline\hline
\end{tabular*}
\end{center}
\end{table*}

\begin{table*}
\renewcommand{\arraystretch}{1.2}
\caption{Same as in Table \ref{ta2} for the $\Delta$-type resonances included into the model.}
\label{ta3}
\begin{center}
\begin{tabular*}{14.4cm}
{@{\hspace{0.3cm}}c@{\hspace{0.3cm}}|@{\hspace{0.3cm}}c@{\hspace{0.3cm}}|
 @{\hspace{0.3cm}}c@{\hspace{0.3cm}}|@{\hspace{0.3cm}}c@{\hspace{0.3cm}}|
 @{\hspace{0.3cm}}c@{\hspace{0.3cm}}|@{\hspace{0.3cm}}c@{\hspace{0.3cm}}
}
\hline\hline\noalign{\smallskip}
 $\Delta(M_R)J^P$ & $M_R$ & $\Gamma_{tot}$
& $\sqrt{\beta_{\pi\Delta}}\,A_{1/2}$ & $\sqrt{\beta_{\rho N}}\,A_{1/2}$ & $A_{3/2}/A_{1/2}$ \\
  & [MeV] & [MeV] & [10$^{-3}$GeV$^{-1/2}$] & [10$^{-3}$GeV$^{-1/2}$] & \\
\noalign{\smallskip}\hline\hline\noalign{\smallskip}
$\Delta(1600)\frac32^+$ & 1614 & 250 & $-55.7$ & $-19.7$ & 0.66 \\
                    & $-$ & $-$ & $-$ & $-$ & $-$ \\
                    & $1570\pm70$ & $250\pm50$ & $-39.7\pm3.2$ & $-$ &  $0.778\pm0.592$ \\
                    \hline\noalign{\smallskip}
$\Delta(1620)\frac12^-$ & 1615 & 140 & $-7.96$ & $-5.61$ &  \\
                    & 1620 & 150 & $12.0$ & $0$ & \\
                    & $1610\pm20$ & $130\pm20$ & $35.4\pm26.0$  & $-$   &  \\
                    \hline\noalign{\smallskip}
$\Delta(1700)\frac32^-$ & 1718 & 300 & $25.5$ & $28.6$ & $1.22$ \\
                    & $-$ & $-$  & $-$  & $-$  & $-$      \\
                    & $1710\pm20$ & $300\pm80$ & $71.2\pm74.6$  & $-$  &  $1.00\pm0.54$\\
                    \hline\noalign{\smallskip}
$\Delta(1905)\frac52^+$  & 1860 & 330 & $8.72$ & $19.4$ & $-1.59$ \\
                    & $-$ & $-$  & $-$  & $-$  & $-$      \\
                    & $1880^{+30}_{-25}$ & $330^{+70}_{-60}$ & $20.9\pm11.7$ & $-$ &  $-2.04\pm0.01$  \\
                    \hline\noalign{\smallskip}
$\Delta(1910)\frac12^+$  & 1882 & 300 & $26.7$ & $2.08$ &  \\
                    & $-$ & $-$  & $-$  & $-$  &      \\
                    & $1900\pm50$ & $300\pm100$ & $14.1\pm15.1$ & $-$ &  \\
                     \hline\noalign{\smallskip}
$\Delta(1950)\frac72^+$  & 1949 & 285 & $-26.3$ & $17.3$ & 0.87  \\
                    & $-$ & $-$  & $-$  & $-$  & $-$     \\
                    & $1930^{+20}_{-15}$ & $285\pm50$ & $-15.6\pm12.9$ & $-$ &  $1.28\pm0.23$ \\
\noalign{\smallskip}\hline\hline
\end{tabular*}
\end{center}
\end{table*}

Another set of observables, $P_x$ and $P_y^\odot$ contribute exclusively to the distribution
over the azimuthal angle $\phi_{23}^*$ (the angle between reaction and production plane)
and vanish when integrated over this angle. These are asymmetries which due to parity
conservation can only appear in three-body final states. These observables have only been analyzed
with analysis (3) (i.e. proton identification with hit multiplicity and carbon background
subtraction). Therefore, the control of systematic uncertainties is less good than for the other
observables. On the other hand, systematic uncertainties from detector effects are less important
for this observables because for a fixed angle between reaction and production plane the data
are integrated over all azimuthal angles in the laboratory coordinate system so that most systematic 
effects from instrumental asymmetries vanish.

The results for $P_x$ and $P_y^\odot$
are shown in Figs.\,\ref{fig:PX} and \ref{fig:PYC}. Parity conservation requires
that
\begin{equation}
P_x(\phi)=-P_x(2\pi-\phi)\,, \quad P_y^\odot(\phi)=-P_y^\odot(2\pi-\phi)\,,
\end{equation}
which is visible in the two figures (it means the asymmetries are invariant when mirrored
around $2\pi$ and the sign of the ordinate is inverse). 

According to the partial wave expansion of \cite{Fix_12} these two observables and the
beam-helicity asymmetry $I^{\odot}$ \cite{Oberle_13} are determined by interferences of partial
waves of the same parity. Therefore, in contrast to the asymmetries $P_y$ and $P_x^\odot$
discussed above, in this case the waves with opposite parities are added incoherently.
Thus, the information contained in the data for $P_x$, $P_y^\odot$ and $I^{\odot}$ is
complementary to those for the first two asymmetries. As may be seen, these observables have
relatively large values only at low incident photon  energies.
With increasing energy the oscillations become smaller. There is no indication of oscillations
with higher frequency (which would be characteristic for contributions from larger angular momenta).

As a theoretical basis we use an isobar model similar to that presented in Ref.\,\cite{Fix_05}.
The reaction amplitude consists of two main terms. The first one includes the nucleon and
$\Delta$-nucleon Born diagrams ((a) to (k) in Fig.\,\ref{fig:diag}). The second term, represented
by the diagrams (l) and (m), contains the sum of $s$-channel Breit-Wigner resonances. In the
calculation presented in Ref.\,\cite{Fix_05} the resonance parameters, including $\gamma N$
coupling and partial decay widths, were taken from the Particle Data Group (PDG) compilation of
Ref.\,\cite{PDG_00}.

In contrast to \cite{Fix_05}, in the present version of the model the resonance parameters were fitted
to the data. Furthermore, besides the Born diagrams in Fig.\,\ref{fig:diag} we included additional 
background terms in the amplitude which were not contained in the model of \cite{Fix_05}.
This takes into account the results of the analysis \cite{Kashevarov_12}.
There, it was found that the amplitude should contain rather large fractions of the partial waves
with $J^P=3/2^\pm$, which are not reproduced by the previous analysis \cite{Fix_05}. As is shown in
\cite{Kashevarov_12}, the background terms seem to be responsible for the steep rise of the total
cross section for $\gamma p\to \pi^0\pi^0 p$ in the region below the first maximum at $W=1500$\,MeV
(see Fig.\,\ref{fig:tot_hyd}). The major constraint of the theory is that these terms should have
a smooth energy dependence.

The resulting model amplitudes, including additional background terms, were fitted to the available
data for $\gamma N\to \pi\pi N$ in all charge channels in the region up to the total energy $W=1900$\,MeV.
The asymmetries $P_x$, $P_y$, $P_x^\odot$, and $P_y^\odot$ obtained in the present work were included
in the fitting procedure. The $N$ and $\Delta$ resonances fitted to the available 
$\gamma N\rightarrow \pi\pi N$ data, along with their resonance parameters are listed in
Tables~\ref{ta2} and \ref{ta3}. 
For the fitting procedure the initial values of the resonance parameters were taken from the 
current Particle Data Group listing \cite{PDG_18}. Since the resonance terms are proportional to the 
product of the electromagnetic and hadronic couplings in the tables the products 
$\sqrt{\beta_\alpha}A_{1/2}$ with $\alpha=\{\pi\Delta,\,\rho N,\,\sigma N\}$ and the ratio of 
the helicity amplitudes $A_{3/2}/A_{1/2}$ is given. The corresponding total cross section as well as 
the target and the beam-target asymmetries are shown in Figs.~\ref{fig:tot_hyd},\ref{fig:Tasym}-\ref{fig:PYC} 
as black solid lines.

In comparison to the analysis of \cite{Fix_05} the new isobar model includes a larger number of resonances, 
especially in the high-mass range. All states with masses up to 1950~MeV and an overall status 
of $****$ were included, as well as the baryon $N(1700)3/2^-$ which according to \cite{PDG_18} has a 
status of $***$ in the $\pi\Delta$ channel. Comparing the results for the states which were included in 
\cite{Fix_05} and the present fit, one observes significant changes to their parameters. To some extent this 
variation is caused by the difference between their values in \cite{PDG_00} and \cite{PDG_18}. This is 
particularly the case for the $\rho N$ and $\sigma N$ decay modes.

The data for $P_y$ and $P_x^\odot$ reveal some interesting properties which should be discussed in
more detail. Firstly, the large values of these asymmetries in the region below 1 GeV are notable, especially 
at energies below the $N(1520)3/2^-$ resonance. This effect is particularly pronounced for the asymmetries 
as function of the invariant masses of the $\pi^0\pi^0$ and $p\pi^0$ pairs. As can be shown by evaluation 
of $P_y$ and $P_{x}^{\odot}$ (using, for example, the partial wave expansion of the t-matrix in 
\cite{Fix_12}), their invariant mass distributions, after integration over the rest of the variables, 
are determined by the interference of partial waves with opposite parities. Therefore, the experimental 
results demonstrate that already at low incident photon energies contributions from both parities are 
important.

The interference effect between the states with opposite parities is also responsible for the
asymmetry in the angular distributions of both observables (see Figs.\,\ref{fig:Tasym} and
\ref{fig:Fasym}). It may be shown, using again the partial wave expansion of \cite{Fix_12},
that when only states of equal parity contribute, the asymmetries $P_y(\Theta_p)$ and
$P_x^\odot(\Theta_p)$ are odd functions of their arguments, so that, for instance,
$P_y(\pi-\Theta_p)=-P_y(\Theta_p)$. This is not true for the measured data
(see Figs.~\ref{fig:Tasym},\ref{fig:Fasym}) which is further evidence that even at low
incident photon energies amplitudes with both parities must contribute.

\section{Summary and conclusions}\label{conclusion}

The present work reports experimental results for the target and beam-target
asymmetry for photoproduction of $\pi^0$ pairs off protons. The data were measured with
a circularly polarized photon beam at the tagged-photon facility of the Mainz MAMI accelerator
and a transversally polarized solid butanol target. The reaction products ($\pi^0$ decay photons
and recoil protons) were detected with the electromagnetic calorimeter combining the Crystal Ball
and the TAPS detectors, supplemented by detectors for charged particle identification.
The results represent a further piece in the puzzle to disentangle the partial wave content
of the $\gamma p\rightarrow p\pi^0\pi^0$ reaction in the second and third resonance regions.

The experimental data are compared to the results of an isobar model with non-resonant background.
The model of Ref.\,\cite{Fix_05} disagrees significantly with the data in the threshold region.
In particular the total cross section above and below the $N(1520)3/2^-$ maximum is not reproduced,
obviously significant contributions are missing. In the range of photon energies between 0.8 and 1~GeV,
in between the double-bump structure in the total cross section, all asymmetries deviate
significantly between the model \cite{Fix_05} and experiment.

The present data for the target and the beam-target asymmetries have been included in a new analysis
of $\gamma N\to\pi\pi N$ at the energies from the threshold to $W=1900$ MeV. Compared to the results
of Ref.\,\cite{Fix_05}, the new solution provides a much better description of the data in the $\pi^0\pi^0p$
channel in the entire energy range.

The measurement of further observables for this reaction is certainly necessary. Some observables
(e.g. reactions with a circularly polarized photon beam and longitudinally
polarized target for quasifree protons and neutrons) have already been measured and are in preparation
for publication.

\section*{Acknowledgments}

We wish to acknowledge the outstanding support of the accelerator group and operators of MAMI.
This work was supported by Schweizerischer Nationalfonds (200020-156983, 132799, 121781, 117601),
Deutsche For\-schungs\-ge\-mein\-schaft (SFB 443, SFB 1044, SFB/TR16), the INFN-Italy,
the European Community-Research Infrastructure Activity under FP7 programme (Hadron Physics,
grant agreement No. 227431), the UK Science and Technology Facilities Council
(ST/J000175/1, ST/G008604/1, ST/G008582/1,ST/J00006X/1, and ST/L00478X/1),
the Natural Sciences and Engineering Research Council (NSERC, FRN: SAPPJ-2015-00023), Canada.
A.F. acknowledges additional support from the Tomsk Polytechnic University Competitiveness 
Enhancement Program.
This material is based upon work also supported by the U.S. Department of Energy,
Office of Science, Office of Nuclear Physics Research Division, under
Award Numbers DE-FG02-99-ER41110, DE-FG02-88ER40415, DE-FG02-01-ER41194,
and DE-SC0014323 and by the National Science Foundation, under
Grant Nos. PHY-1039130 and IIA-1358175.


\end{document}